\documentclass[12pt]{article}
\usepackage{color,amssymb,amsmath,hyperref}
\usepackage{empheq}
\usepackage{url}

\definecolor{rougef}{rgb}{0.56,0,0}
\definecolor{vertf}{rgb}{0,0.5,0}
\definecolor{bleuf}{rgb}{0,0,0.8}

 \csname
@addtoreset\endcsname{equation}{section}

\def\a{\alpha}

\def\b{\beta}  
\def\g{\gamma} 
\def\G{\Gamma}
\def\d{\delta} 

\def\e{\epsilon} 
\def\ve{\varepsilon}
\def\f{\phi}
\def\F{\Phi}

\def\l{\lambda}

\def\r{\rho}
\def\s{\sigma}
\def\S{\Sigma}

\def\q{\theta}

\def\O{\Omega}
\def\o{\omega}
\def\P{\Pi}
\def\U{\Upsilon}
\def\c{\chi}

\def\qq{\quad\quad}

\def\pa{\partial}
\def\na{\nabla}

\def\nn{\nonumber}

\newcommand{\w}[1]{\\[0.#1cm]}
 
\def\be{\begin{equation}}
\def\ee{\end{equation}}
\def\bea{\begin{eqnarray}}
\def\eea{\end{eqnarray}}
\def\ba{\begin{array}}
\def\ea{\end{array}}

\def\w{\wedge}
\def\wt{\widetilde}
\def\ed{\end{document}}

\def\Real{{\mathbb R}}

\begin{document}

\begin{titlepage}

\setcounter{page}{1}

\begin{center}

\hfill

\vskip 1cm


{\LARGE Differential Poisson Sigma Models with }\\[10pt] {\LARGE Extended Supersymmetry}

\vskip 25pt

{\sc Cesar Arias\,$^a$, Per Sundell\,$^b$ and  Alexander Torres-Gomez\,$^c$\\}

\vskip 15pt

{\em $^{a,b}$ \hskip -.1truecm Departamento de Ciencias F\'isicas, 
  Universidad Andres Bello,\\ Republica 220, Santiago de Chile}

\vskip 10pt

{\em $^{c}$ \hskip -.1truecm Departamento de Matem\'aticas, 
  Universidad Cat\'olica del Norte, \\ Avda. Angamos 0610, Casilla 1280 Antofagasta, Chile}

\end{center}

\vskip 15pt

\begin{center} {\bf Abstract}\\[1ex]
\end{center}

\noindent The induced two-dimensional topological $\mathcal N=1$ supersymmetric 
sigma model on a differential Poisson manifold $M$ presented in
\href{http://arxiv.org/abs/arXiv:1503.05625}{arXiv:1503.05625}
is shown to be a special case of the induced Poisson sigma model 
on the bi-graded supermanifold $T[0,1]M$.
The bi-degree comprises the standard $\mathbb N$-valued 
target space degree, corresponding to the form degree on 
the worldsheet, and an additional $\mathbb Z$-valued 
fermion number, corresponding to the degree in the 
differential graded algebra of forms on $M$.
The $\mathcal N=1$ supersymmetry stems from the 
compatibility between the (extended) differential Poisson bracket 
and the de Rham differential on $M$.
The latter is mapped to a nilpotent vector field $\mathcal Q$ 
of bi-degree $(0,1)$ on $T^\ast[1,0](T[0,1]M)$, and 
the covariant Hamiltonian action is $\mathcal Q$-exact.
New extended supersymmetries arise as inner derivatives 
along special bosonic Killing vectors on $M$ that induce 
Killing supervector fields of bi-degree $(0,-1)$ on 
$T^\ast[1,0](T[0,1]M)$.

\vskip 50pt

\begin{flushleft} \footnotesize
{${}^a$ \href{mailto:ce.arias@uandresbello.edu}{ce.arias@uandresbello.edu}}\\
{${}^b$ \href{mailto:per.anders.sundell@gmail.com}{per.anders.sundell@gmail.com}}\\
{${}^c$ \href{mailto:alexander.torres.gomez@gmail.com}{alexander.torres.gomez@gmail.com}}\\
\end{flushleft}

\end{titlepage}


\newpage

{
\tableofcontents }

\vspace{1 cm}

\section{Introduction}
\label{Section:1}

A differential Poisson manifold \cite{Chu,Beggs,Tagliaferro,Zumino} 
is a Poisson manifold whose bracket admits an extension from the algebra 
of functions to the differential graded algebra of forms.
An ordinary Poisson structure induces a two-dimensional 
topological sigma model, due to Ikeda, Schaller and Strobl 
\cite{Ikeda,Schaller}, for which Cattaneo and Felder \cite{Cattaneo} 
devised an Alexandrov-Kontsevich-Schwarz-Starobinsky \cite{AKSZ}
path integral quantization scheme that reproduces Kontsevich's 
star product formula in trivial topology \cite{Kontsevich}.
As shown in \cite{us}, a differential Poisson structure 
induces a fermionic extension of the Ikeda--Schaller--Strobl 
model, referred to as the differential Poisson sigma model,
which couples to the Poisson bi-vector as well as a compatible 
connection\footnote{A differential Poisson manifold
also contains an additional tensorial one-form,
referred to in \cite{us} as the $S$-structure.},
and that has a rigid supersymmetry corresponding to the target 
space de Rham differential.

It remains to be seen whether the path integral quantization 
of the model yields a covariant star product for differential 
forms\footnote{On general grounds, the covariant star 
product must be gauge equivalent in the zero-form sector 
to Kontsevich's canonical product.} that is compatible 
with a suitable deformation of the de Rham differential.
Indeed, McCurdy and Zumino \cite{Zumino} have provided a covariant 
deformation of the wedge product along the differential Poisson 
bracket that is associative but incompatible with the 
undeformed de Rham differential at order $\hbar^2$,
suggesting that compatibility requires a deformed 
differential\footnote{Deformed differentials play
an important role in Fedosov's construction 
of covariant star products of functions on 
symplectic manifolds \cite{Fedosov:1994zz}.}.
It may turn out, however, that the quantization
on general background will actually violate 
associativity up to homotopies, as is often
the case with BRST-like symmetries in field
theory.

As the model's local degrees of freedom 
are confined to boundaries, it describes 
quantum mechanical particles on the Poisson
manifold that are entangled via topological 
fields inside the worldsheet.  
The Poisson sigma model was originally 
developed in response to deformation quantum mechanics 
\cite{Moyal:1949sk,Bayen:1977ha,Bayen:1977hb},
in which the physical state of a quantum mechanical system 
is represented by a density matrix obeying an evolution
equation.
The prospect of the Liouville equation being a linearization of
a string field equation based on a suitably gauged Poisson sigma model, 
which might lead to a microscopic description of quantum 
entanglement and dynamically generated collapses of quantum states, 
provides a basic physical motivation behind the current work.
In other words, topological strings not only implement the 
correspondence principle, whereby classical functions 
are deformed into operators, but also provide a nonlinear
quantum mechanical evolution equation.

To the above end, one needs to distinguish between the gauging 
of Killing symmetries with \cite{Zucchini} and without
Hamiltonians.
The latter category contains the rigid supersymmetry,
which corresponds to the de Rham differential that is 
supposed to give the kinetic term in the (nonlinear) Liouville equation.
As for its gauging, there are two methods available: 
Treating supersymmetry as a fermionic Killing vector in 
target space, it can be gauged at the level of the classical action, 
as was done in \cite{us2}.
Alternatively, treating the supersymmetry as the $Q$-structure of
an integrable $QP$-structure, which means that it can be extended 
naturally to homotopy Poisson manifolds, requires the full 
AKSZ machinery, and we hope to report on it elsewhere.

A related motivation is the proposal made in \cite{Engquist:2005yt} 
that topological open strings and related chiral
Wess--Zumino--Witten models on Dirac cones \cite{Engquist2} are dual
to conformal field theories.
Compared to earlier holography proposals 
\cite{Sundborg:2000wp,Sezgin:2002rt,Klebanov:2002ja},
in which the bulk side is a (string) field theory containing 
higher spin gravity of Vasiliev type \cite{Vasiliev1988,Vasiliev1990},
the proposal of \cite{Engquist:2005yt} was based on the 
observation that tensionless strings in anti-de Sitter 
spacetime behave as collections of conformal particles, 
that is, on a first-quantized description of the bulk 
physics.
Subsequent tests at the level of on-shell amplitudes in $AdS_4$, 
have shown that traces of Gaussian density matrices of the
the quantum mechanical harmonic oscillator in two dimensions
describing unfolded boundary-to-bulk propagators of massless fields 
\cite{Giombi:2010vg}, which one may think of as vertex operators on 
conformal particle worldlines in phase space\footnote{See equation Eq. (3.161) 
in \cite{Engquist:2005yt}.}, or, boundary vertex operators on the corresponding
topological open string, indeed reproduce the correlation 
functions for bilinear operators in free conformal field 
theories in three dimensions \cite{Colombo1, Colombo2, Didenko1, Didenko2}
\footnote{The on-shell correspondence takes the form $\langle O_1 ... O_n \rangle_{\rm CFT} 
= \sum_{\rm crossings} \langle V_1 ... V_n \rangle_{\rm PSM}$, where $O_i$ is a CFT 
operator at a point $x_i$ with conformal label $L_i$ 
and $V_i$ is a Gaussian PSM vertex operator of the form $V_i = \exp( Y^T A_i
Y + Y^T B_i)$ where tbe matrix $A_i$ depends on $x_i$ and $B_i$ depends on $L_i$. 
As the PSM is topological, the right-hand side does not depend on the 
insertion points of the $V_i$ (along the boundary of a disc).}.
The current, slightly refined, working hypothesis is that 
there exists a variant of Witten's realization of Chern-Simons 
theory as a topological A model \cite{Witten1} (see also \cite{Bershadsky}),
whereby a gauging of the differential Poisson sigma model
with 3-graded Chan-Paton factors \cite{Marcus} (see also
\cite{GZ1,GZ2}) yields the Frobenius-Chern-Simons formulation \cite{FCS,FCS2} 
of higher spin gravity.

In this paper, we shall induce the supersymmetric sigma model on 
a differential Poisson manifold using a geometric approach that
yields a covariant Hamiltonian action with canonical kinetic terms.
The method also applies to extended differential Poisson manifolds,
whose brackets contain components with strictly positive intrinsic ç
form degree.
We shall use the resulting manifest target space superdiffeomorphism 
covariance\footnote{The manifest superdiffeomorphism 
covariance is also useful in gauging Killing supersymmetries,
as we hope to report on in a future work.} to finding new extended supersymmetries corresponding to 
inner derivatives of the differential graded algebra of forms
along special bosonic Killing vectors.

The paper is organized as follows: In Section \ref{Section:2} 
we review the induced sigma models on ordinary and differential 
Poisson manifolds. 
In Section \ref{Section:3} we give the geometric construction 
of the supercovariant Hamiltonian action on the phase space 
$T^\ast[1,0](T[0,1]M)$ of an extended differential Poisson 
manifold $M$.
In particular, the geometric origin of the new Killing 
supersymmetries of negative degree is explained in 
Section \ref{sec:newsusy}.
In Section \ref{Section:4} we show the equivalence to the original 
action in the unextended case with vanishing $S$-structure, and 
give the component form of the new Killing supersymmetries.
We conclude and point to future directions for investigations
in Section \ref{Section:5}.
Appendix \ref{Conventions} contains conventions and notation for 
affine connections, the Nijenhuis--Schouten bracket and parity
shifted bundles.
Appendix \ref{App:B} spells out the extension of the Cartan algebra 
to vector field valued differential forms on a real manifold $M$, 
which is mapped in Section \ref{Sec:3.1} to the Cartan algebra on 
$T[1,0]M$. 
Finally, in Appendix \ref{App:C} we provide a manifestly
${\rm Diff}(M)$ covariant formula for the cohomology of 
the Lie derivative of nilpotent vector field ${\cal Q}$
on $T^\ast[1,0](T[0,1]M)$, which uplift representing to
$T^\ast[1,0](T[0,1]M)$ of the de Rham differential on $M$.

\section{Induced two-dimensional Poisson sigma models}
\label{Section:2}

In this section, we review the interplay between
(differential) Poisson algebras and two-dimensional
(supersymmetric) Poisson sigma models.

\subsection{Poisson algebra of functions}
\label{Section:2.1}

The Poisson algebra of functions on a real smooth manifold
$M$ is the space $C^{\infty}(M)$ equipped with its pointwise 
product, which turns it into a commutative associative algebra,
and a second bilinear antisymmetric product 
\be
\label{PB0}
\{\cdot, \cdot\} : C^{\infty}(M) \otimes C^{\infty}(M)
                  \longrightarrow C^{\infty}(M) \ ,
\ee
referred to as the Poisson bracket, acting as a differential 
in each slot and obeying the Jacobi identity. That is, if
$f,g,h\in C^{\infty}(M) $ then
\begin{align}
\label{bilinearity}
&\{f, g+h \} = \{f, g\} + \{f, h\} ~,\\
\label{antisymmetry}
&\{f,g \} = -\{g,f\} ~,  \\
\label{Leibniz}
&\{f,gh\} = g\{f,h\} + h \{f,g\} ~, \\
\label{Jacobi}
&\{f,\{g,h\} \} + \{ h , \{f,g\} \} + \{ g , \{h,f\} \}=0 ~.
\end{align}
The Poisson bracket corresponds to an antisymmetric 
bivector field $\Pi$ defined by
\be \Pi(df,dg) := \tfrac12 \{f,g\}\ ,\ee
obeying the Poisson bivector condition
\be
\{\Pi,\Pi\}_{\rm S.N.}=0\ ,
\ee
where $\{\cdot,\cdot\}_{\rm SN}$ denotes the Schouten--Nijenhuis 
bracket for antisymmetric polyvector fields,
which is equivalent to the Jacobi identity (\ref{Jacobi}).
In a coordinate basis, we may expand
\be \Pi=\tfrac{1}{2}\,\P^{\a\b} \pa_\a \w \pa_\b\ ,\ee
so that
\be
\label{PB}
\{f, g \}= \P^{\a\b} \pa_\a f \, \pa_\b g \ ,
\ee
and 
\be
\{\Pi,\Pi\}_{\rm S.N.}=\P^{\d [\a} \pa_\d \P^{\b\g]} \; \partial_\alpha \wedge \partial_\beta \wedge \partial_\gamma \ .
\ee
Thus, the Poisson bivector condition reads
\be
\label{PoissonCondition}
\P^{\d [\a} \pa_\d \P^{\b\g]} = 0 ~.
\ee

\subsection{Bosonic sigma model}
\label{Section:2.2}

A Poisson manifold $M$ induces a two-dimensional topological field theory 
with configuration space given by the space of maps 
\be \varphi:\Sigma \rightarrow M\ ,\ee
sending a two-dimensional compact surface 
$\Sigma$, with or without boundary, into $M$.
Its classical dynamics is governed by
Ikeda-Schaller-Strobl action \cite{Ikeda,Schaller}
(see also \cite{Cattaneo})
\begin{align}
\label{bPSM}
S[\f, \eta]&=\int_{\Sigma} \varphi^\ast\left( \eta_\a d\f^\a
+ \tfrac{1}{2}\, \P^{\a\b} \, \eta_\a \eta_\b \right) \notag \\
&=\int_{\Sigma} \bigg[ (\varphi^\ast\eta_\a) \w d(\varphi^\ast\f^\a)
+ \tfrac{1}{2}\, (\varphi^\ast\P^{\a\b}) \, (\varphi^\ast\eta_\a) \w (\varphi^\ast\eta_\b) \bigg]
\end{align}
where $\f^\a$ and $\eta_\a$ coordinatize the base and 
fiber of the parity shifted cotangent bundle\footnote{The 
space $T^\ast[1]M$ is obtained from $T^\ast M$ by replacing 
its fiber, that is, the real vector space $\Real^{m}$, 
where $m={\rm dim}(M)$, by the vector space $\Real[1]^{m}$ 
consisting of $m$-tuples of elements in the space $\Real[1]$ 
of real numbers of parity one.
In general, on $\mathbb Z^k$-graded manifold $M$ with
coordinates $Z^i$ with degrees $\overrightarrow{\rm deg}\, Z^i\in \mathbb Z^k$,
the bundles $T[\vec m]M$ and $T^\ast[\vec n]M$ are obtained from
$TM:= T[\vec 0]M$ and $T^\ast M :=T^\ast[\vec 0 ]M$, respectively,
by replacing their fibers by vector spaces of the same dimension
with vectorial coordinates $V^i$ and $P_i$ with $\mathbb Z^k$-valued
degrees given by $\overrightarrow{\rm deg}\, V^i= \overrightarrow{\rm deg}\, Z^i+ \vec m$ 
(shifts) and $\overrightarrow{\rm deg} \,P_i = \vec n - \overrightarrow{\rm deg}\, Z^i$
(dual shifts).
In the bosonic and supersymmetric Poisson sigma models formulated
on Poisson and differential Poisson manifolds, respectively, we have
$k=1$ and $k=2$.}
%
$T^\ast[1]M$ over $M$.
The space $\Omega(T^\ast[1]M)$ of forms on $T^\ast[1]M$ is a differential
graded algebra whose elements have a bi-degree given by the standard form 
degree and a degree (preserved by $\varphi$) dictated by
\be {\rm deg}(\phi^\a,\eta_\a;d)=(0,1;1)\ .\ee
Likewise, the space $\Omega(\Sigma)$ of forms on $\Sigma$ is a differential
graded algebra with degree map given by the form degree on $\Sigma$.
The sigma model map is assumed to have vanishing intrinsic degree 
in the sense that
\be \varphi^\ast:\Omega(T^\ast[1]M) \rightarrow \Omega(\Sigma)\ ,\ee
is a degree preserving homeomorphism of differential graded algebras,
\emph{i.e.} a $p$-form of degree $n$ on $T^\ast[1]M$ is sent to an 
$n$-form on $\Sigma$ and $\varphi^\ast d = d \varphi^\ast$; in particular
\be
 \varphi^\ast (T^\ast [1] M) =\varphi^\ast(T^\ast M) \otimes T^\ast\Sigma ~.\label{eq214}
\ee
The action $S$ is well-defined provided that $(\varphi^\ast\phi^\a,\varphi^\ast\eta_\a)$
belong to a globally defined section over $\Sigma$, which is what we shall assume henceforth,
together with suppressing the symbol $\varphi^\ast$ whenever no ambiguity can arise.

By virtue of the Poisson condition \eqref{PoissonCondition},
the action (\ref{bPSM}) is invariant under the gauge
transformations
\begin{align}
\d_{\ve} \f^\a &= -\P^{\a\b} \ve_\b  ~, \\
\d_{\ve} \eta_\a&= d \ve_\a +  \pa_\a \P^{\b\g} \eta_\b \ve_\g ~,
\end{align}
provided that $\varepsilon^\a$ is globally defined on $\Sigma$, 
and the equations of motion, \emph{viz.}
\begin{align}
\label{eom1}
\cal R^{\a} &:= d\f^\a +  \P^{\a\b} \eta_\b \approx 0 ~,\\
\label{eom2}
\cal R_{\a} &:= d \eta_\a + \tfrac{1}{2} \, \pa_\a \P^{\b\g} \eta_\b \w \eta_\g \approx 0 ~,
\end{align}
define a universally Cartan integrable system,
\emph{i.e.} $d \mathcal R^{\a} \approx 0\,$ and
$d \mathcal R_{\a} \approx 0$ independently of 
the dimension of $\Sigma$.
More precisely, the gauge invariance of the action
and the integrability of \eqref{eom1} are equivalent 
to \eqref{PoissonCondition}, while the integrability 
of \eqref{eom2} only requires the derivative of 
\eqref{PoissonCondition}.

Finally, if $\Sigma$ has a boundary then it is also assumed that
\be
\eta_{\a}|_{{}_{\pa \S}} = 0 ~,\qquad \varepsilon_\a|_{{}_{\pa \S}} = 0\ ,
\ee
as to ensure gauge invariance and that the action is stationary
on-shell including boundary terms.
Strictly speaking, in the classical theory, the variation principle 
only implies that the boundary condition on $\eta_\a$ must hold
on-shell, whereas it must be imposed off-shell in the path integral 
in order for the AKSZ master action to obey the BV master equation.

\subsection{Poisson algebra of differential forms}
\label{Section:2.3}

A differential Poisson algebra on a manifold
$M$ \cite{Chu, Beggs, Tagliaferro, Zumino} is
an extension of a Poisson algebra on $M$ from $C^\infty(M)$ 
to the algebra $\Omega(M)$ of differential forms
on $M$, whereby the pointwise product
on $C^\infty(M)$ is replaced by the graded commutative wedge product 
on $\Omega(M)$, and the Poisson bracket on $C^\infty(M)$
is extended to a graded skew-symmetric map
\be
\label{diffPB}
\{\cdot, \cdot \} : \O(M) \otimes \O(M) \longrightarrow  \O(M) ~,
\ee
referred to as differential Poisson bracket, 
assumed to be compatible with the de Rham differential
and obeying the graded Leibniz rule, that is
\begin{align}
\text{deg}_M (\{ \o_1, \o_2\}) &= \text{deg}_M(\o_1) + \text{deg}_M(\o_2) ~, \label{P1}\\
\{ \o_1, \o_2+\o_3\}&=\{ \o_1, \o_2\} + \{ \o_1, \o_3\} ~,  \label{P2}\\
\{ \o_1, \o_2\} &=(-1)^{1+ \text{deg}_M (\o_1)\text{deg}_M(\o_2)}\{ \o_2, \o_1\} ~,\label{P3}\\
d \{\o_1,\o_2\} &= \{d\o_1,\o_2\} + (-1)^{\text{deg}_M(\o_1)} \{\o_1,d\o_2\} ~,\label{P4}\\
\{ \o_1, \o_2 \wedge \o_3 \} &=\{ \o_1, \o_2\} \wedge \o_3 +
(-1)^{\text{deg}_M(\o_1)\text{deg}_M(\o_2)}\o_2 \wedge \{ \o_1,  \o_3 \} ~,\label{P5}
\end{align}
and that satisfy the graded Jacobi identity
\begin{align}
\label{Jacobis1}
\{ \o_1  ,  \{ \o_2,\o_3 \}  \} &+
(-1)^{\text{deg}_M(\o_1)(\text{deg}_M(\o_2)+\text{deg}_M(\o_3))}\{  \o_2 ,  \{\o_3 , \o_1\}  \} \nonumber\\[5pt]
&+ (-1)^{\text{deg}_M(\o_3)(\text{deg}_M(\o_1)+\text{deg}_M(\o_2))}\{ \o_3  ,  \{\o_1 ,\o_2 \}  \} =0\ ,
\end{align}
where $\omega_i\in \O(M)$ and $\text{deg}_M$ denotes the form degree on $\O(M)$.

Besides a Poisson bi-vector, a differential Poisson bracket entails 
a connection one-form $\widetilde \G^\a{}_\b=d\phi^\g\,\G^\a_{\g\b}$
and a tensorial one-form
\be
\label{Stensor}
S=\tfrac{1}{2}\,
d\phi^\a S^{\b\g}_\a \pa_\b \odot \pa_\g\ ,
\ee
defined through\footnote{Here and in what follows,
we made use of the notation and conventions for affine
connections given in Appendix \ref{Conventions}.}
\be
\{\f^\a, d\f^\b\} = \tfrac{1}{2} \wt \na \P^{\a\b} + S^{\a\b} - \P^{\a\g} \wt \G^{\b}{}_{\g}\ .
\ee
Choosing the connection such that\footnote{Such a connection always exists, 
as there exists a natural generalization of Darboux's theorem from symplectic 
to Poisson manifolds, but it is not unique, leading to a symmetry whereby
the differential Poisson bracket is left invariant under suitable shifts in the 
connection and the tensorial structure $S$, that can be used to eliminate 
$S$ in the symplectic case \cite{us}.}
\be
\wt \na_{\g} \P^{\a\b} = 0 \ ,
\ee
and imposing the compatibility between the bracket and the de Rham differential,
it follows that, for any two differential forms $\omega_1$ and $\omega_2$, one has
\begin{align}
\label{diffPB}
\{\omega_1, \omega_2\} = \Pi^{\a\b} \nabla_{\a} \omega_1 \wedge \nabla_{\b} \omega_2
&+  S^{\a\b} \wedge \left[ (-1)^{\text{deg}_M(\omega_1)} \nabla_{\a} \omega_1 \wedge i_\b \omega_2
- i_\a \omega_1 \wedge \nabla_{\b} \omega_2 \right] \cr
&+ (-1)^{\text{deg}_M(\omega_1)} \left(\widetilde R^{\a\b}
-\widetilde\nabla S^{\a\b} \right) \wedge i_\a \omega_1 \wedge i_\b \omega_2 ~,
\end{align}
where $\nabla_\a$ is constructed from the connection coefficients 
\be \G^\a_{\g\b} = \wt \G^\a_{\b\g}\ ,\ee
which implies 
\be \nabla_\a \Pi^{\b\gamma}= \widetilde \nabla_\alpha \Pi^{\beta
\gamma}-2T^{[\b}_{\a\d} \Pi^{\gamma]\d}\ ,\ee
and where we have defined 
\be
\wt R^{\a\b}:= \P^{\b \g} \wt R^\a{}_\g = \wt R^{\b\a}
\ee
where $\wt R^\a{}_\b$ is the curvature two-form
of $\widetilde \G^\a{}_\b$.
Turning to the graded Jacobi identity \eqref{Jacobis1}, using the properties of the differential Poisson bracket, it holds for all $\omega_i$ if it holds
for ${\rm deg}_M (\omega_1, \omega_2, \omega_3)\in\{(0,0,0),(0,0,1),(0,1,1)\}$. If the tensorial 
structure $S=0$, these conditions are equivalent to
\begin{align}
\label{J000}
{}J_{(0,0,0)}^{\a\b\g}&:=\P^{\d [\a}T^\b_{\d\e} \Pi^{\g]\e} =  0   ~,  \\[5pt]
\label{J001}
{}J_{(0,0,1)}^{\a\b,\g}{}_{\d}&:=\P^{\a \r} \P^{\s \b} R_{\r \s}{}^{\g}{}_{\d}= 0 ~, \\[5pt]
\label{J011}
{}J_{(0,1,1)}^{\a,\b\g}{}_{\d\e}&:=\P^{\a \l} \na_\l \wt R_{\d\e}{}^{\b \g} = 0 ~, 
\end{align}
of which the constraints on $J_{(0,0,0)}^{\a\b\g}$, $J_{(0,0,1)}^{\a(\b,\g)}{}_{\d}$
and $J_{(0,1,1)}^{[\a,\b]\g}{}_{\d\e}$ are independent, whereas the remainder
follows by covariant differentiation.
Moreover, the exterior derivative of the graded Jacobi identity for 
${\rm deg}_M (\omega_1, \omega_2, \omega_3)=(0,1,1)$ yields that for ${\rm deg}_M (\omega_1, \omega_2, \omega_3)=(1,1,1)$,
which reads
\be 
\label{J111}
{}J_{(1,1,1)}^{\a\b\g}{}_{\d\e\l}:=\wt R_{\e [ \r}{}^{(\a \b} \wt R_{\s \l]}{}^{\g)\e} = 0\ .
\ee
A basic example \cite{Beggs} consists of the algebra of functions on a Lie group,
which is deformed, at the quantum level, by the canonical Poisson 
bi-vector into the group algebra, and yet further by the connection 
into the quantum group algebra, for which Eq. 
\eqref{J111} provides the Yang-Baxter equation \cite{Zumino}.

\subsection{$\mathcal N=1$ supersymmetric Poisson sigma model}
\label{Section:2.4}

A differential Poisson manifold induces an $\mathcal N=1$ 
supersymmetric extension of the Ikeda--Schaller--Strobl 
sigma model \cite{us}, obtained by adding fermionic 
partners $(\theta^\a, \chi_\a)$ of form degrees zero 
and one, respectively, to the original bosonic fields 
in the action \eqref{bPSM}.
The classical action is given by
\begin{align}
\label{fPSM}
S[\phi, \eta,\theta,\chi]&= 
           \int_{\Sigma}\varphi^\ast \left( \eta_\a  d\f^\a
           +\tfrac12 \Pi^{\a\b} \eta_\a  \eta_\b
           +\chi_\a \na \theta^\a
           + \tfrac14 \wt R_{\g\d}{}^{\a\b} \c_\a \c_\b \theta^\g\theta^\d \right) \notag \\[5pt]
           &= 
           \int_{\Sigma}\left[  (\varphi^\ast \eta_\a) \w d(\varphi^\ast \f^\a)
           +\tfrac12 (\varphi^\ast \Pi^{\a\b}) (\varphi^\ast \eta_\a) \w (\varphi^\ast \eta_\b)
           + (\varphi^\ast \chi_\a) \w \na (\varphi^\ast \theta^\a) \phantom{\frac 12}\right. \notag \\
           &\qquad \qquad \left.\phantom{\frac 12}+ \tfrac14 (\varphi^\ast \wt R_{\g\d}{}^{\a\b}) (\varphi^\ast \c_\a) \w (\varphi^\ast \c_\b) (\varphi^\ast \theta^\g) (\varphi^\ast \theta^\d) \right]\ ,
\end{align}
where the objects are assigned target space degrees in $\mathbb N$,
denoted by ${\rm deg}$, and fermion numbers in $\mathbb Z$, denoted by 
$n_{\rm f}$, as follows:
\begin{equation}
\begin{tabular}{| l |c | c | c | c | c |} \hline
 &  $\phi^\alpha$ & $\theta^\alpha$ & $ \eta_ \alpha$ & $\chi_\alpha$  & $d$\\ \hline
 $\text{deg}$ &  0 & 0 & 1 & 1 & 1 \\ \hline
 $n_{\rm f}$ & 0 & 1 & 0 & -1 & 0\\ \hline
\end{tabular}
\label{table1}
\end{equation}
The target space thus consists of a bi-graded
fiber bundle 
\be {\cal E}=T^\ast[1,0](T[0,1]M)\equiv T^\ast[1,0]M \oplus T^\ast[1,1]M\oplus T[0,1]M\ ,\ee
with base and fiber coordinatized by $\f^\a$ and $(\eta_\a,\chi_\a,\theta^\a)$,
respectively.
The sigma model map $\varphi:\Sigma\rightarrow M$ is now assumed to
be bi-degree preserving, where the bi-degree on $\Sigma$ is given by
$({\rm deg}_\Sigma,n_{\rm f})$, that is, the pull back operation
$\varphi^\ast$ converts target space degree into form degree on $\Sigma$
(just as in the case of the bosonic sigma model) and preserves the
fermion number\footnote{Put into equations, $\varphi^\ast : \Real[m,n] \rightarrow
\Real[0,n] \otimes \Omega_{[m]}(\Sigma)$, that is, $\varphi^\ast$ converts the 
first entry of the bi-degree into form
degree on $\Sigma$ while preserving the second entry,
given by the fermion number.
}.
Thus, $\varphi^\ast$ induces the following
bundle structure over $\Sigma$:
\begin{align}
\varphi^\ast\Big(T^\ast[1,0]M \oplus T^\ast[1,1]M\oplus T[0,1]M\Big)
      =& \Big(\varphi^\ast(T^\ast[0,0]M)\otimes T^\ast \Sigma\Big) \\
\oplus &\Big(\varphi^\ast(T^\ast[0,1]M)\otimes T^\ast \Sigma \Big)\oplus
\Big(\varphi^\ast(T[0,1]M)\otimes C^{\infty}(\Sigma)\Big) ~, \nonumber
\end{align}
whose sections we shall assume are globally defined on $\Sigma$.
Moreover, in case $\Sigma$ has a boundary, then we assume that
\be \eta_\a|_{\partial \Sigma}=0\ ,\qquad
\chi_\a|_{\partial \Sigma}=0\ .\ee
The kinetic term for the fermions contains the covariant derivative
\be \na \theta^\a:= d\theta^\a+d\phi^\b \Gamma^\a_{\b\g}\theta^\g\ .\label{nablatheta}\ee
The resulting symplectic potential  $\vartheta$ on $\mathcal E$ is given by the sum of
the tautological one-form on $\mathcal E$ and an extra term containing the connection, \emph{viz.}
\be \vartheta=(\eta_\a -\Gamma^\g_{\a\b} \chi_\g \theta^\b) d\phi^\a+\chi_\a d\theta^\a\ .
\label{vartheta}\ee
Finally, we use the following Koszul sign convention\footnote{\label{newK}This convention, which differs from that used in \cite{us}, is motivated by the
fact that it admits a direct extension to generalized Poisson 
sigma models in higher dimensions with target spaces
given by $\mathbb N$-graded manifolds \cite{AKSZ,Park,Ikeda2012}.}:
\be
{\cal F}{\cal F}'=(-1)^{|\cal F| |\cal F ' |} {\cal F}'{\cal F}\ ,\qquad
|{\cal F}| := {\rm deg}({\cal F})+n_{\rm f}({\cal F})
\label{Koszul}
\ee
where ${\cal F}$ and ${\cal F}'$ are functions of
$(\phi^\a,\theta^\a;\eta_\a,\chi_\a)$. We refer to
$|\cal F|$ as the total degree of $\cal F$.

The equations of motion of the
supersymmetric Poisson sigma model
read\footnote{Some signs in Eqs. \eqref{Rphi}--\eqref{Rchi}
differ from those in the corresponding equations
in \cite{us}, where a different Koszul sign 
convention was adopted.}
\begin{align}
\label{Rphi}
\mathcal R^{\f^\a} &:= d\f^\a+ \Pi^{\a\b}\eta_\b \approx 0 ~, \\
\label{Rtheta}
\mathcal R^{\theta^\a} &:= \na \theta^\a
    + \tfrac12 \wt R_{\g\d}{}^{\a\b}\c_\b\theta^\g \theta^\d \approx 0 ~,\\
\label{Reta}
\mathcal R^{\eta_\a}&:= \na \eta_\a
  + R_{\a \b}{}^{\g}{}_{\d}\, d\phi^\b \w \c_\g \theta^\d
 +\tfrac{1}{4}  \na_\a \wt R_{\b\g}{}^{\d\e}\c_\d \w \c_\e  \q^\b\q^\g \approx 0 ~,\\
\label{Rchi}
\mathcal R^{\c_\a} &:= \na \c_\a
                + \tfrac12  \wt R_{\a\d}{}^{\b\g}\c_\b  \w \c_\g \theta^\d \approx 0 ~,
\end{align}
after a dueful suppression of the sigma model map, and where
we use \eqref{nablatheta}, \emph{idem} for $\na \eta_\a$ and $\na \chi_\a$.
These equations form a universally
Cartan integrable system by virtue
of (\ref{J000})-(\ref{J111}), \emph{i.e.}
provided that the differential Poisson
bracket (\ref{diffPB}) obeys the graded
Jacobi identity.
These identities also ensure the gauge
invariance of the action\footnote{The fact that
the symplectic potential \eqref{vartheta} 
is non-canonical implies that
the off-shell gauge transformations differ from the
on-shell ones by terms
proportional to the Cartan curvatures \cite{us}.}, up to boundary
terms that vanish provided the gauge
parameters vanish at the boundary of
$\Sigma$.

Under the isomorphism 
\be \label{iso} C^\infty(T[0,1]M)\cong \Omega(M)\ ,\ee
the de Rham operator on $M$ is sent to
the nilpotent fermionic vector field
\be q_{\rm f}:=\theta^\a \partial_\a\ ,\ee
on $T[0,1]M$.
As found in \cite{us}, its action on $(\phi^\a,\theta^\a)$
can be extended into a (rigid) nilpotent supersymmetry $\delta_{\rm f}$ of 
the action (\ref{fPSM}), given by\footnote{The coefficient 
four-fermi coupling in \eqref{fPSM} is fixed by the rigid supersymmetry 
but not the local symmetries.
}
\begin{align}
\label{susy0}
\d_{\rm f} \f^\a &= \q^\a ~,  \cr
\d_{\rm f} \q^\q &= 0 ~, \cr
\d_{\rm f} \eta_\a &=
\tfrac{1}{2} \wt R_{\b\g}{}^{\d}{}_{\a} \, \c_\d \,\q^\b \q^\g - \G^\g_{\a\b} \, \eta_\g \, \q^\b  ~,\cr
\d_{\rm f}\c_\a &= -\eta_\a + \G^{\g}_{\a\b} \, \c_\g \, \q^\b \ .
\end{align}
This symmetry can be made manifest by writing
\be
\label{V}
S=\d_{\rm f} \int_\Sigma  {\cal V} ~,\qquad 
{\cal V}= -\c_\a \w \left( d\f^\a+\tfrac 1 2 \P^{\a \b} \eta_\b \right)  ~,
\ee
including total derivatives.

\section{Superspace formulation}
\label{Section:3}

In this section, we use the isomorphism between 
the differential form algebra on $M$ and the algebra of 
functions on $T[0,1]M$ to map the differential Poisson 
bracket on $M$ to a supersymmetric Hamiltonian zero-form on 
$T^\ast[1,0](T[0,1]M)$.
The pullback of it and the tautological one-form 
to the worldsheet induces a two-dimensional ${\cal N}=1$ 
supersymmetric topological sigma model.
This geometric approach can be extended to manifolds 
equipped with Poisson brackets that have components
with positive intrinsic form degree.
It also permits a uniform treatment of Killing supersymmetries, 
including the rigid supersymmetry generated by the de Rham 
differential on $M$ as well as new supersymmetries generated
by inner derivatives along ordinary bosonic Killing vectors
on $TM$.

\subsection{Mapping forms on $M$ to functions on $T[0,1]M$ }
\label{Sec:3.1}

The construction of the supersymmetric Poisson sigma model 
in the previous section makes use of the isomorphism 
\eqref{iso} of differential graded algebras, \emph{i.e.}
the bijection
\be V: \Omega_{[n]}(M) \rightarrow \Omega_{[0|(0,n)]}(T[0,1]M)\ ,\label{isoV}\ee
sending $n$-forms $\omega$ on $M$ to functions
$V(\omega)\equiv V_\omega$ on $T[0,1]M$ that are $n$th order
in the fiber coordinates\footnote{The pull-backs $\varphi^\ast V_\omega$
to $\Sigma$ by the sigma model map $\varphi$ are vertex
operators of form degree zero on $\Sigma$.} while 
preserving the associative algebra structure, 
and intertwining the de Rham differential on $M$ with 
a nilpotent vector field $q_{\rm f}$ on $T[0,1]M$, \emph{viz.}
\be V(\omega_1\wedge \omega_2)= V_{\o_1} V_{\o_2}\ ,\qquad  
V \circ d|_M = q_{\rm f} \circ V\ .\label{Vint} \ee
On $T[0,1]M$, the bi-degrees of $q_{\rm f}$ and the de Rham differential 
are given by
\be {\rm bideg}(q_{\rm f})=(0,1)\ ,\qquad {\rm bideg}(d|_{T[0,1]M})=(1,0)\ .\ee
Correspondingly, the algebra of forms on $T[0,1]M$ decomposes
as follows:
\be \Omega(T[0,1]M)=\bigoplus_{n,p} \Omega_{(p,n)}(T[0,1]M)\ ,\qquad
{\rm bideg}(\Omega_{(p,n)}(T[0,1]M))=(p,n)\ ,\ee
where $\Omega_{(p,n)}(T[0,1]M)$ thus consists of $p$-forms on
$T[0,1]M$ that are $n$th order in the fiber coordinates and
their line elements.

The isomorphism $V$ intertwines the derivations of $\Omega(M)$, 
\emph{i.e.} combinations $\imath_\Xi+{\cal L}_{\Xi'}$ of inner and Lie
derivatives along vector field valued forms $\Xi$ and $\Xi'$ on 
$M$ (see Appendix B), with those of $C^\infty(T[0,1]M)$, \emph{i.e.}
the vector fields on $T[0,1]M$.
The induced linear map\footnote{An isomorphism $V:A\rightarrow \widetilde A$ 
between associative algebras induces the isomorphism $V:{\rm Der}(A)\rightarrow {\rm Der}(\widetilde A)$
defined by $V(\delta(a))= V(\delta) V(a)$ for all $a\in A$ and derivations
$\delta\in {\rm Der}(A)$.} 
\be V: {\rm Der}(\Omega(M))\rightarrow  \Gamma(T[0,1]M,T(T[0,1]M))\ ,\ee
is characterized by 
\be V \circ \imath_\Xi=V(\Xi) \circ V\ ,\qquad  V(\omega \wedge \Xi)=V_\omega V(\Xi)\ ,\ee
from which it follows via the Cartan relation ${\cal L}_{\Xi'}=[d,\imath_{\Xi'}]$
and \eqref{Vint} that  
\be V\circ {\cal L}_{\Xi'}= [q_{\rm f},V(\Xi') ] \circ V\ .\label{Xif}\ee

\subsection{Extended differential Poisson bracket} 

The isomorphism also intertwines \emph{extended} differential Poisson brackets $\{\cdot,\cdot\}$ 
on $M$, which by their definition obey \eqref{P1}--\eqref{Jacobis1} and
\be {\rm deg}_M(\{\cdot,\cdot\})=0\quad \mbox{mod $2$}\ ,\ee
with Poisson superbrackets $\{\cdot,\cdot\}_{\rm f}$ on
$T[0,1]N$ that are compatible with $q_{\rm f}$ and
have intrinsic bi-degrees
\be {\rm bideg}(\{\cdot,\cdot\}_{\rm f})=(0,0)\quad \mbox{mod $(0,2)$}\ .\ee
In other words,
\be
 V\circ \{\omega,\eta\}= \{V_\omega, V_\eta\}_{\rm f}\ ,\qquad
 \{V_1,V_2\}_{\rm f}\equiv \Pi_{\rm f}(dV_1,dV_2)\ ,
\label{PBT[0,1]}
\ee
for $V_i\in C^\infty(T[0,1]M)$, where the 
Poisson bi-supervector field $\Pi_{\rm f}$ on $T[0,1]M$ 
obeys 
\be \{\Pi_{\rm f},\Pi_{\rm f}\}_{\rm S.N.} =0\ ,\qquad
{\cal L}_{q_{\rm f}} \Pi_{\rm f}=0\ ,\qquad 
{\rm bideg}(\Pi_{\rm f})=(-2,0)\quad \mbox{mod $(0,2)$}\ ,
\label{Jac}\ee
using the Schouten--Nijenhuis superbracket $\{\cdot,\cdot\}_{\rm S.N.}$ 
for graded antisymmetric polysupervector fields on $T[0,1]M$.

\subsection{Covariant Hamiltonian action on $T^\ast[1,0](T[0,1]M)$}

The Schouten--Nijenhuis superbracket $\{\cdot,\cdot\}_{\rm S.N.}$ 
on $T[0,1]M$ can be mapped to the canonical Poisson bracket 
$\{\cdot,\cdot\}_{(-1,0)}$ on the parity shifted cotangent bundle\footnote{
This map is a classical counterpart of a map used in
the AKSZ procedure \cite{AKSZ} for constructing covariant Hamiltonian 
BV master actions from integrable
polyvector field structures; see also  \cite{Park,Ikeda2012}.}
\be T^\ast[1,0](T[0,1]M)\equiv \left(\Real^m[1,0]\oplus \Real^m[1,-1]\right) 
\hookrightarrow {\cal E}\stackrel{\pi}{\rightarrow } T[0,1]M\ ,\ee
where $m={\rm dim} M$.
To this end, we start from the tautological one-form $\vartheta$ 
on $T^\ast[1,0](T[0,1]M)$, which obeys\footnote{On the total 
space ${\cal E}$ of $T^\ast[1,0](T[0,1]M)$, the space
of forms $\Omega({\cal E})=\bigoplus_{m,n,p}\Omega_{[p|(m,n)]}({\cal E})$
where ${\rm deg}_{\cal E} (\Omega_{[p|(m,n)]}({\cal E}))=p$ and
${\rm bideg}(\Omega_{[p|(m,n)]}({\cal E}))=(m,n)$.}
\be {\rm bideg}(\vartheta)=(2,0)\ ,\qquad \imath_v \vartheta =0 \quad 
\mbox{for all $v\in \text{ker} \,\pi_\ast$}\ ,\label{taut}
\ee
where $\pi$ is the projection map of $T^\ast[1,0](T[0,1]M)$.
The canonical two-form 
\be {\cal O}:= d\vartheta\ ,\qquad {\rm bideg}({\cal O})=(3,0)\ ,
\ee
as ${\rm bideg}(d|_{\cal E})=(1,0)$.
The resulting canonical bracket on ${\cal E}$ has intrinsic bi-degree 
\be {\rm bideg}(\{\cdot,\cdot\}_{(-1,0)})=(-1,0)\ ,\ee
and is graded antisymmetric and obey the graded Leibniz'
rule and Jacobi identity, \emph{viz.}
\begin{align} \{{\cal F}_1,{\cal F}_2\}_{(-1,0)}&=(-1)^{1+(|{\cal F}_1|+1)(|{\cal F}_2|+1)}
\{{\cal F}_2,{\cal F}_1\}_{(-1,0)}\ ,\\
\{{\cal F}_1,{\cal F}_2{\cal F}_3\}_{(-1,0)}&=\{{\cal F}_1,{\cal F}_2\}_{(-1,0)}{\cal F}_3
+(-1)^{|{\cal F}_2|(|{\cal F}_1|+1)}{\cal F}_2\{{\cal F}_1,{\cal F}_3\}_{(-1,0)}\ ,\end{align}
and 
\begin{align}
\{\{{{\cal F}}_1,{{\cal F}_2}\}_{(-1,0)},{{\cal F}_3}\}_{(-1,0)}&+(-1)^{({{\cal F}}_2+
{{\cal F}_3})({{\cal F}}_1+1)}\{\{{{\cal F}_2},{{\cal F}_3}\}_{(-1,0)},{{\cal F}_1}\}_{(-1,0)}\\
&+(-1)^{({{\cal F}_1}+{{\cal F}_2})({{\cal F}_3}+1)}\{\{{{\cal F}_3},{{\cal F}_1}\}_{(-1,0)},{{\cal F}_2}\}_{(-1,0)}= 0\ ,
\end{align}
where ${\cal F}_i$ are functions on ${\cal E}$.
These can be expanded as 
\be {\cal F}=\sum_{n=0}^\infty {\cal F}^{(n)}\ ,\qquad {\cal F}^{(n)}=P^{(n)}(\vartheta^{\wedge n})\ ,\qquad
{\rm bideg}(P^{(n)})={\rm bideg}({\cal F}^{(n)})-(n,0)\ ,\ee
where $P^{(n)}$ are rank $n$ graded antisymmetric polyvector fields 
on $T^\ast[1,0](T[0,1]M)$ defining equivalence classes 
\be \left[P^{(n)}\right]= \left[P^{(n)} + P^{\prime (n-1)}\wedge v\right]\ ,
\qquad v\in {\rm ker}\, \pi_\ast\ ,\label{class}\ee
in view of \eqref{taut}, 
whose projections to the base define (distinct) graded antisymmetric rank $n$ polysupervector fields 
\be P^{(n)}_{\rm f}=\pi_\ast P^{(n)}\ ,\ee
on $T[0,1]M$.
Conversely, one has a bijective uplift $\rho$ as follows:
\be \left[ P^{(n)}\right]=\rho(P^{(n)}_{\rm f})\ ,\qquad 
\pi_\ast\circ \rho={\rm id}\ ,
\label{faithful}\ee
which thus has ${\rm ker} \rho=0$.
Writing $\left[ P^{(n)}\right](\vartheta^{\wedge n})
\equiv P^{(n)}(\vartheta^{\wedge n})$, the relation between 
the canonical Poisson bracket on ${\cal E}$ and the 
Schouten--Nijenhuis superbracket $\{\cdot,\cdot\}_{\rm S.N.}$ 
on $T[0,1]M$ takes the following form:
\begin{align}
\{{\cal F}^{(n_1)}_1,{\cal F}^{(n_2)}_2\}_{(-1,0)} &\equiv
\{\rho(P^{(n_1)}_{{\rm f},1})(\vartheta^{\wedge n_1}),\rho(P^{(n_2)}_{{\rm f},2})(\vartheta^{\wedge n_2})\}_{(-1,0)}\nn\\[5pt]
&=\rho(\{P^{(n_1)}_{{\rm f},1},P^{(n_2)}_{{\rm f},2}\}_{\rm S.N})(\vartheta^{\wedge (n_1+n_2-1)})\ .
\end{align}
Next, the vector field $q_{\rm f}$ on $T[0,1]M$ is uplifted
to a nilpotent vector field ${\cal Q}$ on ${\cal E}$ defined by
\be 
\label{Qtheta}
{\cal L}_{\cal Q}\vartheta=0\ , \qquad
\pi_\ast {\cal Q}=q_{\rm f}\ ,\qquad {\rm bideg}({\cal Q})=(0,1)\ .
\ee
Thus, from $\pi_\ast ({\cal L}_{\cal Q} \rho(P^{(n)}_{\rm f}))={\cal L}_{q_{\rm f}} P^{(n)}_{\rm f}$
and \eqref{faithful} it follows that  
\begin{align} {\cal L}_{\cal Q}  {\cal F}^{(n)}&\equiv {\cal L}_{\cal Q}\left( \rho(P^{(n)}_{\rm f})(\vartheta^{\odot n})\right)\cr
&\equiv ({\cal L}_{\cal Q} \rho(P^{(n)}_{\rm f}))(\vartheta^{\odot n})+(-1)^{|{\cal F}^{(n)}|}n
 \rho(P^{(n)}_{\rm f}))(({\cal L}_{\cal Q}\vartheta)\odot \vartheta^{\odot (n-1)})\cr
 &=\rho({\cal L}_{q_{\rm f}} P^{(n)}_{\rm f})(\vartheta^{\odot n})\ .\end{align}
 
Turning to the supersymmetric Hamiltonian function 
$\cal H$ on $\cal E$, by definition it obeys\footnote{The condition 
on the bi-degree implies that ${\cal H}$ is quadratic in momenta.
In the AKSZ approach, this implies that  ${\cal H}$ vanishes 
on the trivial section as required by the boundary conditions
following from the BV master equation.} 
\be
\label{QH} 
{\cal Q}  {\cal H}=0\ ,\qquad \{{\cal H},{\cal H}\}_{(-1,0)}=0\ ,\qquad 
{\rm bideg}({\cal H})=(2,0)\ \mod\ (0,2)\ ,\ee
from which it follows that  
\be {\cal H}= \rho(\Pi^{(2)}_{\rm f})(\vartheta^{\wedge 2})\ ,\ee
where $\Pi^{(2)}_{\rm f}$ obeys \eqref{Jac}.
Furthermore, as demonstrated in Appendix \ref{App:C},
locally on ${\cal E}$ there exist a one-form $\cal G$ and a function
${\cal W}$ such that 
\be
\vartheta = {\cal L}_{\cal Q} {\cal G} ~, \qquad {\cal H} = {\cal Q} {\cal W}\ ,\qquad
{\rm bideg}({\cal G})={\rm bideg}({\cal W})=(2,-1)\ .
\ee
The resulting covariant Hamiltonian action of the classical theory reads
\be S=\int_{\Sigma}\varphi^\ast(\vartheta+{\cal H})=\delta_{\rm f} \int_{\Sigma}\varphi^\ast {\cal V}\ ,\qquad 
{\cal V}:={\cal G}+{\cal W}\ ,\label{S}\ee
where the sigma model map $\varphi:\S \rightarrow T[0,1]M$
is assumed to have vanishing intrinsic bi-degree, \emph{i.e.}
\be {\rm bideg}(\varphi)=(0,0)\ ,\ee
and obey the boundary condition 
\be \varphi: \partial\Sigma\rightarrow T[0,1]M\ ,\ee
that is, the boundary of $\Sigma$ is sent to the trivial 
section on $T^\ast[1,0](T[0,1]M)$, \emph{i.e.} the momenta
vanishes at $\partial \Sigma$.

\subsection{Formulation in local coordinates}

We coordinatize $\mathcal E$ using ($i=0,1$)
\begin{align}  
\Phi^\a_i&=(\Phi^\a_0,\Phi^\a_1)\equiv (\phi^\a,\theta^\a)\ ,\qquad 
\text{bideg}(\Phi^{\a}_{i}) = (0,i)\ ,\\
H_\a^i &= (H_\a^0,H_\a^1)\equiv (\omega_\a,\chi_\a) ,\qquad {\rm bideg}(H^i_\a)=(1,-i)\ .
\end{align}
where $H_\a^i$ and $\Phi^\a_i$ are coordinates of the fiber and the 
base of $T^\ast[1,0](T[0,1]M)$, respectively, and $\phi^\a$ and 
$\theta^\a$ are coordinates of the base and fiber of $T[0,1]M$, respectively.
The Koszul sign convention \eqref{Koszul} yields 
the following graded commutativity relations:
\be
\Phi^{\a}_{i} \Phi^{\b}_{j} = (-1)^{ij} \Phi^{\b}_{j}\Phi^{\a}_{i}  \ ,\quad
\Phi^{\a}_{i} H^{j}_{\b} =  (-1)^{i(1+j)} H^{j}_{\b} \Phi^{\a}_{i} \ , \quad
 H^{i}_{\a} H^{j}_{\b} =  (-1)^{(1+i)(1+j)} H^{j}_{\b} H^{i}_{\a}\ .\label{gcr2}
\ee
The nilpotent vector field $q_{\rm f}$ is given by
\be 
q_{\rm f}=\theta^\a \frac{\partial}{\partial\phi^\a}\equiv q_i{}^{j} \Phi^\a_j \partial_\a^i\ ,\qquad q_i{}^j=\delta_i^0\delta^j_1\ ,\qquad 
\partial_\a^i:=\frac{\partial}{\partial \Phi^\alpha_i}\ .
\ee
As for the canonical two-form and tautological one-form, we take 
\be {\cal O}= dH_\a^i \wedge d \Phi^\alpha_i\ ,
\qquad \vartheta=H_\a^i d\Phi^\a_i\ .\ee
It follows that the uplift ${\cal Q}$ to ${\cal E}$ of $q_{\rm f}$ on $T[0,1]M$ 
is given by
\be {\cal Q}=q_i{}^j (\Phi^\a_j \partial_\a^i - H_\a^i \partial^\a_j)=q_{\rm f}+ \widetilde q_{\rm f}\ ,
\qquad \widetilde q_{\rm f}=-\omega_\a \frac{\partial}{\partial\chi_\a}\ ,\ee
obeying
\be (q_{\rm f})^2=\{q_{\rm f},\widetilde q_{\rm f}\}=(\widetilde q_{\rm f})^2=0\ .\ee
Expanding the Poisson bi-supervector on $T[0,1]M$ as
\be \Pi_{\rm f}= \Pi^{\a\b}_{ij} \partial_\a^i \w \partial_\b^j\ ,\qquad
\partial_\a^i \w \partial_\b^j=-(-1)^{ij} \partial_\b^j \w \partial_\a^i\ ,\ee
where thus
\be \P^{\a\b}_{ij} = -(-1)^{ij}  \P^{\b\a}_{ji}\ ,\qquad
\text{deg}( \Pi^{\a\b}_{ij})=0\ ,\qquad n_{\rm f}( \P^{\a\b}_{ij}) = i+j\ ,\ee
and using ${\rm bideg}(d\Phi^\a_i)=(1,i)$, the coordinate
form of the superbracket \eqref{PBT[0,1]} reads
\be \{f_1,f_2\}_{\rm f}= (-1)^{i+j(|f_1| +1)} \Pi^{\a\b}_{ij} \partial_\a^i f_1
\partial_\b^j f_2\ .\ee
The Poisson bi-supervector and $q_{\rm f}$-compatibility conditions
take the following form:
\begin{align} 
\Pi^{\a\d}_{il} \partial_\d^l \Pi^{\b\g}_{jk} \,\partial_\a^i \w
\partial_\b^j \w \partial_\g^k &=0\ ,\label{qfinv0}\\[5pt]
\left(q_{\rm f}\Pi^{\b\g}_{kl}  +
2 \, \Pi^{\gamma\b}_{lj} q^j_k  \right) \partial_\b^k\w \partial_\g^l &=0\ .
\label{qfinv}
\end{align}
The corresponding supersymmetric 
Hamiltonian function on ${\cal E}$ is given by
\be {\cal H}=\rho(\Pi_{\rm f})(\vartheta\odot \vartheta)=\tfrac12 \Pi^{\a\b}_{ij}H_\a^i  H_\b^j \ .\ee
The explicit form of the resulting covariant Hamiltonian action \eqref{S} takes the form 
\be
\label{hatS}
S=\int_{\S}\varphi^\ast \left(H^{i}_{\a}  d\Phi^{\a}_{i} +\tfrac{1}{2}\,
\Pi^{\a\b}_{ij} H^{i}_{\a}  H^{j}_{\b} \right) \ ,
\ee
where the form degrees on $\Sigma$ and fermion numbers of the pulled back fields are given by

\begin{equation}
\label{table2}
\begin{tabular}
{| l |c | c | c | c | c |} \hline
 &  $\varphi^\ast\Phi^\alpha_0$ & $\varphi^\ast\Phi^\alpha_1$ & $\varphi^\ast H_\a^0$ & $\varphi^\ast H_\a^1$  & $d$\\ \hline
 $\text{deg}_\Sigma$       & 0 & 0 & 1 & 1 & 1 \\ \hline
 $n_{\rm f}$ & 0 & 1 & 0 & -1 & 0\\ \hline
\end{tabular}
\end{equation}
Suppressing $\varphi^\ast$, the equations of motion read
\begin{align}
\label{R1}
\mathcal R^{\a}_i &:=  d\F^{\a}_{i} + (-1)^{i(i+j)} \P^{\a\b}_{ij} H^{j}_{\b} \approx 0 \ ,\\
\label{R2}
\mathcal R_{\a}^i &:= dH^{i}_{\a} + (-1)^{i} \frac{1}{2} \, \pa_\a^i \P^{\b\g}_{jk}\,  H_{\b}^{j} \w H_{\g}^{k}  \approx 0 ~,
\end{align}
which form a universally Cartan integrable system
by virtue of \eqref{qfinv0} and \eqref{qfinv}.
The rigid nilpotent supersymmetry transformation, \emph{viz.}
\be\label{3.47}
\d_{\rm f} \Phi^\a_i = q_i{}^j  \Phi^\a_j  ~~, \qq \d_{\rm f} H_\a^i = - q_j{}^i H_\a^j\ ,
\ee
leaves the action invariant, as follows from  
\be
\label{potential}
S= \d_{\rm f}\int_{\S}  {\cal V} \ ,\qquad
{\cal V}= -H_{\a}^{i} \w  (q^{\rm T})_{i}{}^{j}\left( d\F^\a_{j} + \tfrac{1}{2} \,  \P^{\a\b}_{jk} H_{\b}^{k}  \right)\ ,\qquad(q^{\rm T})_{i}{}^{j}= \delta_i^1\delta^j_0\ ,
\ee
as can be seen by using \eqref{qfinv}, that is,
\be
\label{hdP}
\d_{\rm f} \,  \P^{\a\b}_{ij} = (-1)^{j+k} q_{i}{}^{k}  \P^{\a\b}_{kj} - (-1)^{k} q_{j}{}^{k}
\P^{\a\b}_{ik}\ ,
\ee
or more explicitly
\bea
\d_{\rm f} \,  \P^{\a\b}_{00} = -2\,  \P^{[\a\b]}_{01}\  ,\qquad  \d_{\rm f} \,  \P^{\a\b}_{01} =  \P^{\a\b}_{11}\  ,\qquad
\d_{\rm f} \,  \P^{\a\b}_{10 } =  -\P^{\a\b}_{11}\  ,\qquad \d_{\rm f} \,  \P^{\a\b}_{11} =  0 ~ .
\eea

As can be seen from Tables \eqref{table1} and 
Table \eqref{table2}, the spectra of fields in the
supersymmetric actions \eqref{fPSM} and \eqref{hatS} 
agree.
Indeed, in the case of a differential Poisson manifold, 
there exists a simple field 
redefinition that maps \eqref{fPSM} to \eqref{hatS} 
(without the need to add any total derivative), as will be 
spelled out in detail in Section 4.


\subsection{Rigid supersymmetries from Killing supervectors}\label{sec:newsusy}


The notion of a symmetry of a Poisson algebra of functions
refers to a vector field $K$ on $M$ whose Lie derivative 
annihilates the Poisson bi-vector field, \emph{viz.}
\be {\cal L}_K \Pi=0\ ,\ee
as this is equivalent to that
\begin{align} {\cal L}_K \{f,g\}&\equiv 2{\cal L}_K (\Pi(df,dg))\equiv 2\left(({\cal L}_K\Pi)(df,dg)+
\Pi({\cal L}_K df,g)+\Pi(f,{\cal L}_K dg)\right)\nn\\
&=2\left(\Pi(d{\cal L}_K f,g)+\Pi(f,d{\cal L}_K g)\right)=\{{\cal L}_K f,g\}+\{f,{\cal L}_K g\}\ ,\end{align}
for all $f,g\in C^\infty(M)$.
Such a vector, which is often referred to as a Killing (or fundamental) vector,
induces a rigid symmetry of the Ikeda--Scheller--Strobl sigma model. 

The above notions have natural generalizations to
the context of differential Poisson algebras and 
their induced supersymmetric sigma models.
Thus, a symmetry of an extended differential Poisson algebra
refers to a vector field valued $p$-form $K$ (see
Appendix \ref{App:B}) obeying
\be  {\cal L}_K \{\omega_1,\omega_2\}=\{{\cal L}_K \omega_1,\omega_2\}+(-1)^{{\rm deg}_M(\omega_1)
{\rm deg}_M(K)}\{\omega_1,{\cal L}_K \omega_2\}\ ,\label{KVM}\ee
for all $\omega_1,\omega_2\in \Omega(M)$. 

A general vector field valued $p$-form $\Xi$ on $M$ is sent by
the isomorphism $V$ in \eqref{isoV} to the supervector field $\Xi_{\rm f}=V(\Xi)$
on $T[0,1]M$ defined by \eqref{Xif}, which in its turn induces a supervector field 
${\cal X}_{\Xi}$ on ${\cal E}$ defined by
\be \pi_\ast {\cal X}_{\Xi}=\Xi_{\rm f}\ ,\qquad {\cal L}_{{\cal X}_{\Xi}}\vartheta=0\ ,\qquad
{\rm bideg}({\cal X}_{\Xi})={\rm bideg}(\Xi_{\rm f})\ . \label{uplift}\ee
Thus, in a coordinate basis where $\Xi_{\rm f}=\Xi^\a_i\partial_\a^i$, we have 

\be 
\delta_{\Xi_{\rm f}}\Phi^\a_i\equiv {\cal X}_{\Xi} (\Phi^\a_i)=\Xi_i^\a\ ,\qquad
\delta_{\Xi_{\rm f}} H_\a^i\equiv {\cal X}_{\Xi} (H_\a^i)= -(-1)^{i+j + i |\Xi_{\rm f}|} \partial_\a^i \Xi^\b_j H_\b^j  \ .
\ee
It follows that
\be \delta_{\Xi_{\rm f}} S[\Phi,H;\Pi_{\rm f}]
= {\cal L}_{\Xi_{\rm f}}S[\Phi,H;\Pi_{\rm f}]\ ,\label{fieldvsbg}\ee
where thus $\delta_{\Xi_{\rm f}}$ acts only on the fields and 
${\cal L}_{\Xi_{\rm f}}$ acts only on the background field.

Thus, the vector field valued $p$-form $K$ of Killing type is mapped to
supervector fields $K_{\rm f}=V(K)$ on $T[0,1]M$ and
${\cal X}_{K}$ on ${\cal E}$, obeying
\be {\cal L}_{K_{\rm f}} \Pi_{\rm f}=0\ ,\qquad {\cal X}_{K} {\cal H}=0\ ,\ee
and hence it generates a global symmetry, \emph{viz.}
\be \delta_{K_{\rm f}} S[\Phi,H;\Pi_{\rm f}]=0\ .\ee
In this language, the compatibility between the extended differential
Poisson bracket and the de Rham differential amounts to that the vector 
field valued one-form $I=d\phi^\a \partial_\a$ on $M$ is of Killing type.
It is mapped by $V$ to $q_{\rm f}=V(I)=q_i^j  \Phi^\alpha_j \partial_\a^i$, 
hence inducing the rigid nilpotent supersymmetry transformation \eqref{3.47}.
An ordinary Killing vector field $K=K^\a \partial_\a$ on $M$, on the order hand,
induces the Killing supervector 
\be K_{\rm f} = K^{\a}_i \partial_{\a}^i\ ,\qquad V\circ {\cal L}_K={\cal L}_{K_{\rm f}}\circ V\ ,
\qquad {\rm bideg}(K_{\rm f})=(0,0)\ ,\ee
on $T[0,1]M$ with components given by
\be  
K^\a_0 = K^\a ~~, \qquad  K^\a_1 = -\theta^\b \partial_\b K^\a   ~.
\ee 
The inner derivative of forms on $M$ along an ordinary Killing vector\footnote{Conversely, 
if $K$ is an ordinary vector field on $M$ and $\imath_K$ is a symmetry of the (extended) 
differential Poisson bracket then $K$ must be an ordinary Killing vector field.} 
$K$, which is mapped by $V$ to
\be \widetilde K_{\rm f}= \widetilde K^{\a}_i \partial_{\a}^i\ ,\qquad V\circ \imath_K={\cal L}_{\widetilde K_{\rm f}}\circ V\ ,
\qquad {\rm bideg}(\widetilde K_{\rm f})=(0,-1)\ ,\label{specKV1}\ee
on $T[0,1]M$, is a symmetry as well provided that $\imath_K$ commutes with
the extended Poisson bracket, or equivalently, that
\be {\cal L}_{\widetilde K_{\rm f}}\Pi_{\rm f}=0\ ,\label{specKV}\ee
whose component form will be derived below in Section 4.2 in the 
differential Poisson case.
%

\section{Component formulation}
\label{Section:4}

In this section, we identify the supersymmetric 
model in Section \ref{Section:2} as the special case of 
the model in Section \ref{Section:3} that arises 
on differential Poisson manifolds with vanishing 
$S$-tensor.
We shall also include non-vanishing $S$-tensors, 
and derive the supplementary conditions on a 
Killing vector on $M$ for it to yield an extra supersymmetry 
of bi-degree $(0,-1)$.

\subsection{Action and equations of motion}

In order to obtain the action \eqref{fPSM} from \eqref{hatS}, we take 
\be\label{fieldredef}
\Phi^{\a}_{i} = (\f^{\a}, \q^{\a}) ~,~~~  
H_{\a}^{i} = (\eta_\a - \G_{\a\b}^{\g} \c_\g \q^\b, \c_{\a})\ ,
\ee
where $\G_{\a\b}^{\g}$ are the coefficients of the 
connection one-form of the differential Poisson algebra.
In the unextended case, we have\footnote{
In the extended case, the Lagrangian contains 
additional terms of bi-degrees $(2,2k)$ for $k\geq 1$.
Fitting these models into the minimal AKSZ geometry 
requires working with actions whose degree vanishes
mod two.}
\begin{align}
\label{abcd}
\P^{\a\b}_{00} &= \P^{\a\b} \cr
\P^{\a\b}_{01} &=-\left(S^{\a\b}_{\g} + \P^{\a\d}\G^{\b}_{\d\g} \right) \q^{\g}\cr
\P^{\a\b}_{10} &= \P^{\b\a}_{01} \cr
\P^{\a\b}_{11} &= \left( \tfrac{1}{2} \, \U_{\g\d}{}^{\a\b}
                  - 2 \,S^{\e(\a}_\g \, \G^{\b)}_{\e\d}
                  + \P^{\e\l} \G^\a_{\e\g} \G^\b_{\l\d} \right) \q^\g\q^\d
\end{align}
where we have separated the combination
\be
\U_{\g\d}{}^{\a\b} := \widetilde R_{\g\d}{}^{\a\b}
                  - 2\, \widetilde \na_{[\g}  S_{\d]}^{\a\b} +  T_{\g\d}^\e S^{\a\b}_\e ~.
\ee
Plugging \eqref{fieldredef} and \eqref{abcd} back into the
action (\ref{hatS}), we obtain
\be
\label{Smodel}
S=\int_{\S} \biggl[ \eta_\a \wedge d\f^\a  + \c_\a \wedge \nabla \q^\a
         + \tfrac{1}{2}\, \P^{\a\b} \eta_\a \wedge \eta_\b
         + S^{\a\b}_{\g} \, \eta_\a \wedge \c_\b \, \q^\g
         +  \tfrac{1}{4} \, \U_{\g\d}{}^{\a\b}  \c_\a  \wedge \c_\b \, \q^\g \q^\d  \biggr] ~,
\ee
which reduces to \eqref{fPSM} upon setting $S^{\a\b}_\g=0$.
Varying the action with respect to  $(\eta, \c, \q)$, respectively,
yields
\begin{align}
\label{Rf}
\cal R^{\f^\a}&:=  d\f^\a + \P^{\a\b}\eta_\b + S_\g^{\a\b} \c_\b \q^\g \approx 0\ , \\
\label{Rq}
\cal R^{\q^\a}&:=  \nabla \q^\a  +  S_\g^{\a\b} \eta_\b \q^\g
                         +\tfrac{1}{2}\, \U_{\g\d}{}^{\a\b} \c_\b \,\q^\g\,\q^\d \approx 0\ ,  \\
 \label{Rc}
 \cal R^{\c_\a}&:=  \nabla \c_\a - S_\a^{\b\g} \eta_\b \wedge \c_\g
             + \tfrac{1}{2}\, \U_{\a\b}{}^{\g\d} \c_\g \wedge \c_\d \,\q^\b  \approx 0\ ,
\end{align}
while its variation with respect to $\f$ yields
\begin{align}
 {\cal R^{\eta_\a}} := d \eta_\a + \tfrac{1}{2}\, \partial_\a  \P^{\b\g} \eta_\b \wedge \eta_\g
    &+ \left( \pa_\a S^{\g\d}_\b \eta_\g \wedge \c_\d - \G^\g_{\a\b} \, d\c_\g
     + 2\, \pa_{[\a} \G^\g_{\d]\b} \c_\g \wedge d\f^\d  \right) \q^\b  \cr
     &- \G^\g_{\a\b} \c_\g \wedge d\q^\b
     + \tfrac{1}{4} \, \pa_\a \U_{\b\g}{}^{\d\e} \c_\d \wedge \c_\e \,  \q^\b\q^\g \approx 0~,
\end{align}
which can be rewritten in a manifestly
covariant form by using the compatibility condition
$\pa_\a \P^{\b\g} = 2\, \G^{[\b}_{\d\a} \P^{\g]\d}$ and
(\ref{Rf})--(\ref{Rc}), with the result
\begin{align}
\label{Reta}
 \mathcal R^{\eta_\a}= \nabla \eta_\a 
 + R_{\a\b}{}^{\g}{}_{\d} \, d\f^\b \wedge \c_\g \, \q^\d
+  \left( \nabla_\a S_\b^{\g\d} - T_{\a\e}^\g S_\b^{\e\d}   \right)   \eta_\g \wedge \c_\d \, \q^\b
+ \tfrac{1}{4} \, \nabla_\a \U_{\b\g}{}^{\d\e} \c_\d \wedge \c_\e \q^\b\q^\g \ .
\end{align}
%

\subsection{Supersymmetries}

\paragraph{${\cal N}=1$ supersymmetry (de Rham operator):} By construction, 
the action  (\ref{Smodel})  can be written on the 
manifestly globally supersymmetric form \eqref{potential}, 
\emph{viz.}
\be
S=\d_{\rm f} \int_{\S}  {\cal V}\ ,\qquad {\cal V}= -\c_\a  \w \left( d\f^\a + \tfrac{1}{2} \, \P^{\a\b} \eta_\b + \tfrac{1}{2} \, S^{\a\b}_\g \c_\b \q^\g  \right)
\ee
where the rigid nilpotent supersymmetry transformation,
which is given in general by \eqref{3.47}, takes the 
following form in terms of the component fields 
$(\phi^\a,\theta^\a;\eta_\a,\chi_\a)$:
\begin{align}
\label{susy1}
\d_{\rm f} \f^\a &= \q^\a  \ ,\cr
\d_{\rm f} \q^\a &= 0  \ ,\cr
\d_{\rm f} \eta_\a &=\tfrac{1}{2} \wt R_{\b\g}{}^{\d}{}_{\a} \, \c_\d \,\q^\b \q^\g - \G^\g_{\a\b} \, \eta_\g \, \q^\b \ , \cr
\d_{\rm f}\c_\a &= -\eta_\a + \G^{\g}_{\a\b} \, \c_\g \, \q^\b\ .
\end{align}
Its nilpotency can be verified using the compatibility condition 
$\widetilde \nabla_\a \P^{\b\g}=0$ and the Bianchi identity  
$\wt \na_{[\a} \wt R_{\b\g]}{}^{\d}{}_{\e} - \wt T^\l_{[\a\b} \wt R_{\g] \l}{}^{\d}{}_{\e} =0$.
%

\paragraph{Extended supersymmetry (inner derivatives):}

Let us demonstrate at the level of components that a Killing 
vector $K=K^\a \pa_\a$ obeying \eqref{specKV} yields an additional 
nilpotent rigid supersymmetry given by \eqref{uplift} and \eqref{specKV1},
\emph{i.e.}
\begin{align}
 \d_{{\rm f},K} \f^\a &= 0   ~, \\
 \d_{{\rm f},K} \q^\a&= K^\a  ~,   \\
 \d_{{\rm f},K} \eta_\a&= \c_\b \na_\a K^\b     ~,  \\
 \d_{{\rm f},K} \c_\a&= 0      ~,
\end{align}
which thus acts non-trivially only on the fields with odd total degree.
Indeed, by \eqref{uplift} the variation of the kinetic 
term under $\d_{{\rm f},K}$, given by the pull back of 
the symplectic potential, vanishes identically (without the need
to use the Killing vector property), whereas the variation 
of the Hamiltonian term reads
\be
\label{tildedS}
\d_{{\rm f},K} S = \int_{\S} \left[  \left( \P^{\g\a} \na_\g K^\b -
K^\g S_\g^{\a\b} \right) \eta_\a \w \c_\b
                + \left( \tfrac{1}{2} \, K^\g \U_{\g\d}{}^{\a\b} +
                S^{\b\g}_\d \na_\g K^\a  \right)  \c_\a \w \c_\b \, \q^\d  \right]\ ,
\ee
that vanishes iff the two terms vanish separately. 
We recall that 
\begin{align}
\label{LG}
\mathcal L_{K} \G^{\g}_{\a\b} & \equiv  \pa_\a \pa_\b K^{\g} - \pa_\a K^\d \G^{\g}_{\d \b}
                              - \pa_\b K^\d \G^{\g}_{\a \d} + \pa_\d K^\g \G^{\d}_{\a\b}=0\ , \\
\label{LP}
\mathcal L_{K} \P^{\a\b} & \equiv  K^\g \pa_\g \P^{\a\b} + 2\, \pa_\g K^{[\a} \P^{\b]\g}=0\ ,  \\
\mathcal L_{K} S^{\a\b}_{\g}& \equiv  K^\d \wt \na_\d S^{\a\b}_\g  +
S^{\a\b}_\d \na_\g K^\d - 2\,S^{\d(\a}_\g\na_\d K^{\b)}=0\ .
\label{LS}
\end{align}
The first equation can equivalently be written as
\be\label{KVid1}
\wt \na_\a \na_\b K^\g = K^\d \wt R_{\a \d}{}^{\g}{}_\bº ,
\ee
and the second equation combined with the compatibility condition
$\pa_\g \P^{\a\b}= 2\, \G^{[\a}_{\d\g} \P^{\b]\d}$ yields
\be
\label{AKcondition}
\P^{\g[\a}\na_\g K^{\b]} = 0 ~.
\ee
Thus, first term in \eqref{tildedS} vanishes iff the following stronger version of
\eqref{AKcondition} holds:
\be
\label{FullKcondition}
\P^{\g\a} \na_\g K^\b = K^\g S_\g^{\a\b}\ .
\ee
As for the second term in \eqref{tildedS}, using the $\wt \na$-derivative of \eqref{FullKcondition}
together with $\wt \na_\g \P^{\a\b}=0$, \eqref{LS} and \eqref{KVid1}, it can be rewritten as
\be
\d_K S = 2\, \int_{\S}  K^\g \wt R_{\g \d}{}^{\a \b} \c_\a  \w \c_\b \, \q^\d  ~,
\ee
whose vanishing requires
\be \label{KRie}
K^\g \wt R_{\g \d}{}^{\a \b} =0\ .
\ee
Thus, in summary, a Killing vector $K$ induces an extended supersymmetry of bidegree
$(0,-1)$ if it obeys the additional conditions \eqref{FullKcondition} and \eqref{KRie},
which can be shown to be equivalent to \eqref{specKV}.
Moreover, from the Cartan relations $\{\imath_K,d\}={\cal L}_K$ and $(\imath_K)^2=0$ it follows that
\be
\{ \d_{{\rm f},K}, \d_{\rm f} \}=\delta_K\ ,\qquad \{\d_K, \d_K\} = 0\ ,
\ee
where $\delta_K$ denotes the action of the ordinary Killing vector $K$ on the
fields, \emph{viz.}
\begin{align}
\delta_K\phi^\a &= K^\a   \cr
\delta_K\theta^\a &=\theta^\b \partial_\beta K^\a  \cr
\delta_K\omega_\a &= \partial_\alpha \partial_\beta K^\gamma \chi_\gamma \theta^\beta
                         - \omega_\beta \,\partial_\alpha K^\beta     \cr
\delta_K\chi_\a &=  - \chi_\beta \partial_\alpha K^\beta \ .    
\end{align}
%

\section{Conclusion and outlook}
\label{Section:5}

We have reformulated the supersymmetric extension \cite{us} 
of the Ikeda--Schaller--Strobl model \cite{Ikeda,Schaller} 
induced on a differential Poisson manifold $M$ as a special
case of the induced sigma model on the bi-graded supermanifold 
$T[0,1]M$ equipped with a Poisson superbracket corresponding to
an extended differential Poisson bracket on the differential
graded algebra of forms on $M$ with intrinsic form degree 
valued in $\{0,2,\dots\}$.
The resulting covariant Hamiltonian action on the phase
space $T^\ast[1,0](T[0,1])M$ is manifestly ${\rm Diff}(T[0,1])$
covariant.
Consequently, it is manifestly invariant under global 
symmetries generated by Killing supervector fields on 
$T[0,1]M$, which we have used to give new extended 
supersymmetries associated to inner derivatives 
along special Killing vectors fields on $M$.

The current model is a special case of the more general sigma model
with canonical action
\be S_{\rm can} = \int_\Sigma \left( H_{\underline\a} \w d\Phi^{\underline\a}
+\tfrac12 {\cal P}^{{\underline\a\b}}  H_{\underline\a} \w  H_{\underline\b}
\right)\ + \oint_{\partial\Sigma} H_{\underline \alpha} {\cal B}^{\underline \alpha}  ,\ee
where ${\cal P}:={\cal P}^{{\underline\a\b}}\partial_{\underline\a}\wedge {\underline\b}$ 
is a Poisson bi-supervector on target supermanifold of type $(m|m')$ 
coordinatized by $\Phi^{\underline\a}=(\Phi^{\a},\Phi^{\a'})$, $\a=1,\dots,m$, 
$\a'=1,\dots,m'$.
The case $m=m'$ is distinguished, however, by the facts that it can 
be made to exhibit the de Rham-like supersymmetry and that it is possible 
to quantize the model using a minimal AKSZ gauge fixing procedure without 
additional trivial pairs (instead of the direct extension of the 
Cattaneo--Felder scheme), as we shall report on in a separate work
\footnote{The resulting BRST differential will have a total degree 
in $\{1,3,\dots\}$ unless ${\cal P}$ is taken to be unextended.}
Whether this is tied to the subtleties of the work of McCurdy and Zumino 
\cite{Zumino} remains to be seen.

Annother advantage of the canonical form of the action is that it facilitates 
the gauging of Killing (super)symmetries using the direct supersymmetrization 
of Zucchini's bosonic formalism \cite{Zucchini}.
Concerning the gauging of the original rigid supersymmetry, a subtlety arises
as there are two approaches available, depending on whether it is gauged 
as a Killing supersymmetry as in \cite{us2}, or if it treated together 
with ${\cal P}$ as an integrable $QP$-structure, as we shall present
in a forthcoming work.

Concerning the perturbative quantization of the model, it remains 
to be investigated whether it yields a differential graded associative 
algebra or if quantum corrections will induce a homotopy associative 
structure. 
As shown in \cite{Beggs}, the former structure arises for special 
differential Poisson geometries related to quantum groups.
However, the results of \cite{Zumino} indicate that on more general 
manifolds\footnote{The analyses of \cite{Beggs} and \cite{Zumino} did not 
include the tensorial one-form $S$ defined in Eq. \eqref{Stensor}.} 
there is an incompatibility between the ``canonical'' associative star 
product and the de Rham differential at order $\hbar^2$, though 
there analysis did not exclude the possibility that compatibility 
can be restored by deforming the differential. 
Formally, the argument that the star product is 
compatible with a deformed differential goes as follows: 
The ${\rm Diff}(T[0,1]M)$ covariance of the classical action is broken
down to ${\rm Diff}(M)$ by means of the minimal gauge fixing procedure 
and the background field method (using covariant derivatives on $M$ 
for the Taylor expansion of the background fields)
Assuming that there exists a generalization of the Cattaneo--Felder 
subtraction scheme that yields an associative binary product map ${\rm mult}_2$,
the action of $\varphi\in {\rm Diff}(T[0,1]M)$ on it is equivalent to a 
Kontsevich-style supergauge transformation $G$, \emph{viz} 
\be ((\varphi)^{-1})^\ast {\rm mult}_2 (\varphi^\ast V_\omega,
\varphi^\ast V_\eta;(\varphi^{-1})_\ast \Pi_{\rm f}) =
G^{-1}{\rm mult}_2 (G V_\omega,
G V_\eta;\Pi_{\rm f})\ ,\ee
using the notation of Section \ref{Section:3} wherein $V_\omega$ is the 
function on $T[0,1]M$ corresponding to the form $\omega$ on $M$.
Since the background is $q_{\rm f}$-invariant it follows that
\be Q_{\rm f}\circ {\rm mult}_2 = {\rm mult}_2\circ (Q_{\rm f}\otimes 1 + 1\otimes Q_{\rm f})\ ,\qquad
Q_{\rm f}=q_{\rm f}+ A_{\rm f}\ ,\ee
where $A_{\rm f}=\sum_{n\geqslant 1} \hbar^n A^{(n)}_{\rm f} $ 
is the multi-differential operator generating the supergauge
transformation induced by $q_{\rm f}$.
Thus, assuning that the subtraction scheme does not yield any anomalies 
in the conservation law for the current of $Q_{\rm f}$, then one has 
the flatness condition
\be Q_{\rm f}\circ Q_{\rm f}=0\ ,\ee
of a differential graded associative algebra.
We leave it for future work to settle the above subtleties
in more detail and whether there will be a need for homotopies 
in the associativity rule. 

Alternatively, higher products can be introduced already at the 
semiclassical level by considering homotopy Poisson algebras given 
by  sets of $n$-ary brackets, for $n\geq 2$, obeying Jacobi identity 
up to homotopies. 
To our best understanding, corresponding induced homotopy Poisson
sigma models have not been studied in the literature, and we plan 
to address them in a future publication.

As pointed out in \cite{FCS2}, quantum homotopy associative algebras 
can be used to extend the cubic Frobenius--Chern--Simons gauge theory 
by employing the tensor constructions of \cite{GZ1,GZ2}.
Thus, drawing further on \cite{Witten1,Bershadsky}, we expect 
that the off-shell formulation of higher spin gravity on 
general backgrounds requires a deformation of the Chern-Simons-like 
cubic action found in \cite{FCS} consisting of simultaneously 
i) adding quadratic and higher order terms to its Hamiltonian function leading to an 
``internal'' homotopy associative algebra generated by generalized 
Chan-Paton-like factors corresponding to ``discrete'' degrees of freedom
of an induced homotopy Poisson sigma model; and ii) replacing the 
differential graded associative algebra of forms on the base manifold 
valued in the higher spin associative algebra by an ``external'' 
homotopy associative algebra corresponding to ``continuous'' degrees
of freedom of the sigma model.
The resulting topological string field would thus be valued in the 
direct product \cite{GZ1,GZ2} of two first-quantized homotopy 
associative algebras.
One may speculate that such topological open string field theories
on target spaces with boundaries may lead to realizations of mirror symmetry 
transformations as nontrivial transitions between topologically inequivalent 
boundary states.

Finally, let us point to a few interesting direction for future research:
First of all, the results that we have accumulated so far are
supportive of the working hypothesis that higher spin gravity 
on a noncommutative manifold $M$ i) is dual to first-quantized open strings; 
ii) has a formulation as a second-quantized topological theory on $M$;
iii) admits a sum over topologies of $M$ that is dual to a third-quantized 
theory, which is supported in part by the fact that the kinetic terms
make up infinite-dimensional abelian $p$-form systems
\cite{Schwarz:1978cn,Schwarz:1984wk,Horowitz:1989km,Wu:1990ci} 
for which there are cancellations \cite{FCS}\footnote{See discussion
of Eqs. (7.1) and (7.2) in \cite{FCS}.} leading 
to a well-defined partition functions at one-loop.
The goal of these investigations is to establish that 
the above types of dualities provide a good ``quantum
gauge principle'' for fundamental interactions in nature.
A related idea, also mentioned in the Introduction, is that
that topological open string fields of suitably gauged models 
contains zero-forms identifiable with density matrices obeying 
nonlinear quantum mechanical evolution equations.


\paragraph{Acknowledgments.}
We are thankful to Severin Barmeier, Roberto Bonezzi, Nicolas Boulanger, 
Rodrigo Canto, Elizabeth Gasparim, Maxim Grigoriev, Kevin Morand, Ergin Sezgin
and Yihao Yin for useful discussions.
C.A. would like to thank the hospitality of 
Groupe de M\'ecanique et Gravitation at Universit\'e de Mons,
where part of this project was carried out.
C.A. is supported by a UNAB PhD scholarship.
The work of P.S. is supported by Fondecyt Regular grant N$^{\rm o}$ 
1140296, Conicyt grant DPI 20140115 and UNAB internal grant DI-1382-16/R.
A.T.G. is thankful to Euihun Joung for the invitation to 
visit the SNU String Theory Group where some part of this work was 
developed. 
A.T.G. was supported by the National Research Foundation of Korea 
through the grant NRF- 2014R1A6A3A04056670 during that visit.
%

\appendix

\section{Conventions and notation}
\label{Conventions}

\paragraph{Affine connections.}

Given a manifold $M$ with tangent bundle $TM$ and tensor bundle
${\cal T}=\bigoplus_{m,n\in \mathbb M} TM^{\otimes m} \otimes T^\ast
M^{\otimes n}$,
which is an associative algebra with product $\otimes$, an affine
connection is a
${\cal C}^\infty(M)$-linear map $\nabla: TM  \rightarrow {\rm
Der}({\cal T})$ that
commutes to the diffeomorphism invariant subalgebra of ${\rm End}({\cal T})$,
which is generated by contractions and insertions of the identity tensor.
The connection $\nabla$ is normalized such that if $X$ is a vector field
and $\phi\in {\cal C}^\infty(M)$ then
\be \nabla_{X} (\phi)= X(\phi )\ ,\ee
using the notation $\nabla_X\equiv \nabla(X)$.
Thus, $\nabla_X: TM^{\otimes m} \otimes T^\ast M^{\otimes n} \rightarrow
TM^{\otimes m} \otimes T^\ast M^{\otimes n} $,
and if $T,T' \in  {\cal T}$ then
\begin{align} \nabla_{\phi X} T&= \phi \nabla_X T\ ,\\
\nabla_{X} (T\otimes T') &= (\nabla_X T)\otimes T' + T\otimes \nabla_X T'\ .
\end{align}
From the compatibility between the connection and the contraction map it
follows that if $V$ is a vector field and $\omega$ is a one-form then
\be \nabla_X (\omega(V))= (\nabla_X \omega)(V)+\omega(\nabla_X V)\
.\label{omegaV}\ee
In terms of local coordinates $\phi^\alpha$, the connection is characterised
by the one-form
\be \Gamma^\a{}_\b=d\phi^\gamma \Gamma^{\a}_{\gamma \b}\ ,\ee
where the connection coefficients are defined by
\be \nabla_{\partial_\alpha} \partial_\beta= \Gamma_{\alpha\b}^\g
\partial_\g\ .\ee
From \eqref{omegaV} and $d\phi^\a (\partial_\beta)=\delta^\a_\b$ it follows that
\be \nabla_{\partial_\alpha} d\phi^\beta= -\Gamma_{\alpha\g}^\b d\phi^\g\ .\ee
For notational simplicity, when acting on a tensor $T= T_{\a \dots }{}^{\b \dots} d\f^\a \otimes \cdots \otimes \pa_\b \otimes \cdots$,
we write
\be
\na_{\pa_\a} T  \equiv \na_\a T \equiv (\na_\a T_{\b \cdots}{}^{\g \cdots})  d\f^\b \otimes \cdots \otimes \pa_\g \otimes \cdots ~,
\ee
where thus
\be
\nabla_{\a} T_{\b \dots}{}^{\g \dots} := \partial_{\a} T_{\b \dots}{}^{\g \dots}
                                         - \G_{\a \b}^{\d} T_{\d \dots}{}^{\g \dots} - \cdots
                                         + \G_{\a\d}^{\g} T_{\b \dots}{}^{\d \dots} + \cdots ~.
\label{covder}
\ee
Given a pair $(X,Y)$ of vector fields, the torsion and Riemann two-forms
$T)\in TN$ and $R\in {\rm End}({\cal T})$, respectively, are defined
by the decomposition
\be [\nabla_X,\nabla_Y] =\nabla_{\nabla_X Y -\nabla_Y X - T(X,Y)}+R(X,Y)\ .\ee
Thus, upon expanding
\be T=\frac12 d\phi^\a \wedge d\phi^\b T_{\a\b}\ ,\qquad
T_{\a\b}=T_{\a\b}^\g \partial_\g\ ,\ee
\be R=\frac12 d\phi^\a \wedge d\phi^\b R_{\a\b}\ ,\qquad
R_{\a\b}\partial_\g=R_{\a\b}{}^\d{}_\g \partial_\d\ ,\ee
it follows from
\be \left[\nabla_{\a} ,
\nabla_{\b}\right]\partial_\g=
\nabla_{\a}(\G^\d_{\b\g} \partial_\d)- (\a\leftrightarrow
\b)\ ,\ee
that
\be
 R_{\a\b}{}^{\gamma}{}_{\d}= 2\, \partial_{[\a} \G_{\b]\d}^{\gamma} +
2\, \G_{[\a|\e}^{\gamma} \G_{|\b]\d}^{\e}  \ ,\qquad
 T_{\a\b}^{\gamma} = 2\, \G_{[\a\b]}^{\gamma}\ .
\ee
Alternatively, in terms of the covariant derivatives defined in
\eqref{covder} one has
\be
[\nabla_{\a}, \nabla_{\b}] V^{\gamma} = - T_{\a\b}^{\d} \nabla_{\d}
V^{\gamma}+R_{\a\b}{}^{\gamma}{}_{\d} V^{\d} \ .\ee
If $T,T'\in {\cal T}$ and $X$ is a vector field, we define
\be \nabla_{T\otimes X} T'= T\otimes \nabla_X T'\ .\ee
It follows that if $\omega$ and $\omega'$ are forms then
\be \nabla_{\omega\otimes X} (\omega'\otimes X')= \omega\wedge (
\nabla_X \omega'\otimes T'+\omega'\otimes  \nabla_X T')\ .\ee
Introducing the vector field valued one-form $I=d\phi^\alpha \partial_\alpha$,
we define, in a slight abuse of notation, the exterior covariant derivative
$\nabla: \Omega(M)\otimes {\cal T} \rightarrow \Omega(M)\otimes {\cal T} $
as follows
\be \nabla (\omega \otimes T)= \nabla_I (\omega\otimes T)\ ,\ee
whose action thus take the following form in components:
\be \nabla (\omega \otimes T)= d\omega \otimes T+d\phi^\alpha
\wedge\omega \otimes \nabla_{\partial_\alpha} T\ .\ee
In particular, acting on a $p$-form $\omega=\tfrac1{p!} d\phi^{\a_1}
\wedge \cdots\wedge
d\phi^{\a_p} \o_{\a_1\dots\a_p}$ one has
\be \nabla\omega=d\omega =  \tfrac1{p!} d\phi^{\a_1}\w \cdots\w
 d\phi^{\a_{p+1}}\left(\nabla_{\a_1} \omega_{\a_2\dots
\a_{p+1}}+\tfrac{p}2 \,T^\b_{\a_1\a_2} \o_{\b\a_3\dots\a_{p+1}}\right)
\ .\ee
Finally, the action of the exterior covariant derivatives of
components is defined via
\be \nabla (V^\alpha \partial_\a) =( \nabla V^\a) \partial_\a\ ,\qquad
\nabla (d\phi^\alpha\omega_\a)=-d\phi^\a \nabla \omega_\a\ ,\ee
where thus
\be  \nabla V^\a=dV^\alpha+\Gamma^\a{}_\b V^\b=d\phi^\b \nabla_\alpha
V^\beta\ ,\qquad
\nabla \omega_\alpha= d\omega_\a-\G^\b{}_\a \omega_\b=d\phi^\b
\nabla_\b \omega_\a\ .\ee
Defining the torsion and curvature two-forms
\be T^\a= \tfrac12 d\phi^\gamma \wedge d\phi^\d T_{\gamma\d}^\a\ ,\qquad
R^\a{}_\b= \tfrac12 d\phi^\gamma \wedge d\phi^\d R_{\gamma\d}{}^\a{}_\b\ ,\ee
which can be rewritten as
\be T^\a= d\phi^\b\wedge d\phi^\g \G^\a_{\b\g}
=\G^\a{}_\b \wedge d\phi^\b= \nabla d\phi^\a  \ ,\qquad R^\a{}_\b=
d\G^\a{}_\b + \G^\a{}_\gamma \wedge \G^\gamma{}_\b\ .\ee
one has the Ricci identities
\be \nabla^2 V^\a = R^\a{}_\b V^\b\ ,\qquad \nabla^2 \omega_\a= -
R^\b{}_\a \wedge \omega_\b\ ,\ee
and the Bianchi identities
\be \nabla T^\a= R^\a{}_\b \wedge d\phi^\b\ ,\qquad \nabla R^\a{}_\b = 0 \ ,\ee
or, in components,
\be
R_{[\a\b}{}^{\gamma}{}_{\d]} = \nabla_{[\a} T_{\b\d]}^{\gamma} -
T_{[\a\b}^{\e} T_{\d] \e}^{\gamma}\ ,\qquad \nabla_{[\a}
R_{\b\gamma]}{}^{\d}{}_\e- T^\lambda_{[\a\b} R_{\gamma]
\lambda}{}^\d{}_\e=0\ .\ee

\paragraph{Polyvector fields and Schouten--Nijenhuis brackets.}

The graded spaces 
\be
\text{Poly}^{(\pm)}(M)=\bigoplus_{n\in \mathbb Z} \text{Poly}^{(\pm)}_{[n]}(M)
\ee
where 
$\text{Poly}^{(\pm)}_{[n]}(M):=0$ for $n\leqslant -2$ and
$\text{Poly}^{(\pm)}_{[-1]}(M):=C^\infty(M)$,
and where for $n\geqslant 0$
\be
\text{Poly}^{(-)}_{[n]}(M):=TM^{\wedge (n+1)}\ , \qquad 
\text{Poly}^{(+)}_{[n]}(M):=TM^{\odot (n+1)}\ ,
\ee
have degree map
${\rm deg}(\text{Poly}^{(\pm)}_{[n]}(M)):=n$ 
and degree preserving Schouten--Nijenhuis brackets 
$\{\cdot,\cdot\}^{(\pm)}_{\rm S.N.}$
defined by 
\be
\{A, B\}^{(\pm )}_{\rm S.N.}=- (\pm 1)^{({\rm deg} (A)+1)({\rm deg} (B)+1)}
\{B, A\}^{(\pm )}_{\rm S.N.}\ ,
\ee 
and obey the Leibniz' rule
\begin{align}  
\{A, B\wedge C\}^{(-)}_{\rm S.N.}&=
\{A , B\}^{(\pm)}_{\rm S.N.} \wedge C+(\pm 1)^{({\rm deg} (A)+1){\rm deg} (B)}
B\wedge  \{B , C\}^{(\pm )}_{\rm S.N.}\ ,\\
\{A, B\odot C\}^{(\pm)}_{\rm S.N.}&=
\{A , B\}^{(+)}_{\rm S.N.} \odot C+
B\odot  \{B , C\}^{(+ )}_{\rm S.N.}\ ,
\end{align} 
where $\{A,B\}^{(\pm )}_{\rm S.N.}:=A(B)$ if ${\rm deg}(A,B)=(0,-1)$
and $\{A,B\}^{(\pm )}_{\rm S.N.}:=[A,B]$ if ${\rm deg}(A,B)=(0,0)$.

\paragraph{Parity shifted bundles.}

We recall that an element $u_p$ of $T^\ast M$ 
over a point $p\in M$ can be expanded as $u_p=u_\a(p)
e^\a_p$ where $u_\a(p)$ are real numbers and $e^\a_p$ 
is a basis for the fiber over $p$, \emph{i.e.} $\pi(e^\a_p)=p$, and that 
a section $u$ of $T^\ast M$ is a map from $M$ to $T^\ast M$ 
such that $u: p \mapsto u_p$, \emph{i.e.} $\pi \circ u = {\rm Id}_M$. 
Moreover, a section $\varphi^\ast u$ of the pulled back bundle $\varphi^\ast T^\ast M$ over $\Sigma$ obeys  
$(\varphi^\ast u)_x = u_{\varphi(x)} = u_\a(\varphi(x)) e^\a_{\varphi(x)}$ for all $x\in\Sigma$.
To apply $\varphi^\ast$ to $T^\ast[n]M$ we first define
$\varphi^\ast(\Real[n]_{\varphi(x)}) \in (T_x^\ast\Sigma)^{\wedge n}$,
that is, a real number $r[n]$ of degree $n$ of a fiber space
over a point $p$ in the image of $\varphi$ that is mapped by $\varphi^\ast$
to an $n$-form on $\Sigma$. 
Thus, as an element of $T^\ast[n]M$ is of the form $u[n]_p
= u_\a[n](p) e^\a_p$ where $u_\a[n](p)\in \Real[n]$ and 
$e^\a_p$ is the basis of the fiber of $T^\ast M$, the 
application of $\varphi^\ast$ to $T^\ast[n]M$ yields 
the right hand side of Eq. \eqref{eq214} for $n=1$.
Alternatively, one can redefine the notation, and take $\varphi:T[1]\Sigma\rightarrow T^\ast[1]M$ 
to be a map of vanishing intrinsic degree, such that $\eta_\alpha\circ \varphi$ is a linear 
function in the fiber coordinate of $T[1]\Sigma$, and use the
canonical definition of $\varphi^\ast:\Omega(T^\ast[1]M)\rightarrow \Omega(T[1]\Sigma)$.
The Lagrangian can then be obtained by composing $\varphi^\ast$ with the homeomorphism
$\mu:\Omega(T[1]\Sigma)\rightarrow \Omega(\Sigma)$ of differential graded algebras,
defined in local coordinates $(x^\mu,\theta^\mu)$ on $T[1]\Sigma$ by 
$\mu:(x^\mu,\theta^\mu;dx^\mu,d\theta^\mu)\mapsto (x^\mu,dx^\mu;dx^\mu,0)$
(such that $\mu\circ d=d\circ \mu$),
\emph{viz.} $S=\int_{\Sigma} \mu\circ \varphi^\ast\left( \eta_\a d\f^\a
+ \tfrac{1}{2}\, \P^{\a\b} \, \eta_\a \eta_\b \right)$.

\section{Graded Lie algebra of derivations of the differential form algebra}
\label{App:B}

The graded Lie algebra ${\rm Der}(\Omega(M))$ of derivations 
of the differential graded associative algebra $\Omega(M)$ 
of forms on $M$ consists of maps 
\be \delta: \Omega{[p]}(M)\rightarrow \Omega{[p+{\rm deg}(\delta)]}(M)\ ,\qquad
{\rm deg}(\delta)\in \{-1,0,1,\dots\}\ ,\ee
obeying the graded Leibniz' rule
\be \delta(\omega\wedge \eta)=(\delta \omega)\w \eta+(-1)^{{\rm deg}(\omega){\rm deg}(\delta)} \omega
\w (\d \eta)\ ,\qquad
\omega, \eta\in \Omega(M)\ .\ee
The graded commutator of $\delta_1,\delta_2\in {\rm Der}(\Omega(M))$ is 
the element in ${\rm Der}(\Omega(M))$ defined by
\be [\delta_1,\delta_2]:=\delta_1\delta_2-(-1)^{{\rm deg}(\delta_1){\rm deg}(\delta_2)}\d_2 \d_1\ .\ee
A general element $\delta\in {\rm Der}(\Omega(M))$ can be decomposed into 
a Lie derivative ${\cal L}_{K}$ and an inner derivative $\imath_{\Xi'}$, \emph{viz.}
\be \delta={\cal L}_{\Xi}+\imath_{\Xi'}\ ,\ee
where $\Xi$ and $\Xi'$ are vector field valued forms on $M$.
If $\Xi=\omega\otimes X$, where $\omega\in \Omega_{[k]}(M)$ and 
$X$ is a vector field on $M$, then we define
\be {\rm deg}(\Xi)=k\ ,\ee
and 
\be \imath_\Xi\eta :=\omega\w (\imath_X \eta)\ ,\qquad {\rm deg}(\imath_\Xi):={\rm deg}(\Xi)-1\ ,\ee
for all $\eta\in \Omega(M)$.
The Lie derivative 
\be {\cal L}_\Xi:=[\imath_\Xi, d]\equiv \imath_\Xi d+ (-1)^{k} d\imath_\Xi \ ,
\qquad {\rm deg}({\cal L}_\Xi)={\rm deg}(\Xi)\ ,\label{defLK}\ee
where $d$ denotes the exterior derivative on $M$, which 
is itself a derivation of $\Omega(M)$ of degree one
that can be represented as a Lie derivative, \emph{viz.}
\be d\equiv {\cal L}_I\ ,\qquad {\rm deg}(d)=1\ ,\ee
where $I$ is the vector field valued one-form defined by
\be \imath_I \omega={\rm deg}(\omega)\omega\ ,\ee
for all $\omega\in \Omega(M)$; in a coordinate basis, we have
\be I=d\phi^\a \partial_\a\ ,\qquad d=d\phi^\a {\cal L}_{\partial_\a}\ .\ee
It follows that exterior and Lie derivatives commute in the graded sense, that is
\be [d,{\cal L}_\Xi]\equiv d {\cal L}_\Xi - (-1)^{{\rm deg}(\Xi)} {\cal L}_\Xi d =0\ .\label{dLbracket}\ee
The inner and Lie derivatives form two subalgebras of ${\rm Der}(\Omega(M))$, \emph{viz.}
\be [\imath_\Xi,\imath_{\Xi'}]=\imath_{[\Xi,\Xi']_{[-1]}}\ ,\qquad [{\cal L}_\Xi,{\cal L}_{\Xi'}]= {\cal L}_{[\Xi,\Xi']_{[0]}}\ ,\ee
while their mutual graded commutator
\be[{\cal L}_\Xi,\imath_{\Xi'}]=\imath_{[\Xi,\Xi']_{[0}}-(-1)^{{\rm deg}(\Xi)({\rm deg}(\Xi')+1)}
{\cal L}_{\imath_{\Xi'}\Xi}\ ,\label{Libracket}\ee
where the induced brackets, whose subscripts indicate their intrinsic degrees,
are the Nijenhuis--Richardson bracket
\be [\Xi,\Xi']_{[-1]}=\imath_\Xi \Xi'-(-1)^{({\rm deg}(\Xi)+1)({\rm deg}(\Xi')+1)} \imath_{\Xi'}\Xi\ ,\ee
where, for $\Xi=\omega\otimes X$ and $\Xi'=\omega'\otimes X'$, we have defined
\be \imath_{\omega\otimes X}\left(\omega'\otimes X'\right):=(\omega\w\imath_X\omega')\otimes X'\ ,\ee
and the Fr\"olicher--Nijenhuis bracket
\begin{align}
[\omega\otimes X,\omega'\otimes X']_{[0]}&=
\omega\wedge\omega'\otimes [X,X']+\left(\omega\w {\cal L}_X \omega'+(-1)^{{\rm deg}(\o)}d\omega\w
\imath_{X}\omega'\right)\otimes X'\nonumber\\[5pt]
&-\left({\cal L}_{X'}\o\w\omega'-(-1)^{{\rm deg}(\o)}
\imath_{X'}\omega\w d\omega'\right)\otimes X\ .
\end{align}
The Nijenhuis--Richardson bracket follows readily,
while the Fr\"olicher--Nijenhuis bracket can be obtained by
combining \eqref{defLK}, \eqref{dLbracket} and \eqref{Libracket} as follows:
\be [{\cal L}_\Xi,{\cal L}_{\Xi'}]=[{\cal L}_\Xi,[\imath_{\Xi'},d]]=[[{\cal L}_\Xi,\imath_{\Xi'}],d]={\cal L}_{[\Xi,\Xi']}\ .\ee
It remains to show \eqref{Libracket}, for which it suffices to
verify that it holds when acting on zero-forms, say $\phi$, and one-forms,
say $\lambda$, for $\Xi=\omega\otimes X$ and $\Xi'=\omega'\otimes X'$ 
with $\omega$ and $\omega'$ being even forms
(after which the general case follows by applying graded degree shifts
to $\Xi$ and $\Xi'$ and making use of the derivation property).
To this end, acting on $\phi$, we have
\begin{align} 
[{\cal L}_{\omega\otimes X},\imath_{\omega'\otimes X'}]\phi&=-\omega'\wedge\imath_{X'}(\omega\imath_X d\phi)\nonumber\\[5pt]
&=-(\omega'\wedge\imath_{X'}\omega)\imath_X d\phi=-\imath_{\omega'\w\imath_{X'}\omega\otimes X}d\phi=-{\cal L}_{\imath_{\omega'\otimes  X'}\omega}\phi\ ,
\end{align}
in immediate agreement with \eqref{Libracket} (as 
$\Xi$ has been assumed to be even).
Acting on $\lambda$ and using $\imath_X\imath_{X'}\l=0$, we have
\begin{align} 
[{\cal L}_{\omega\otimes X},\imath_{\omega'\otimes X'}]\lambda&=d\left(\o\w \imath_X(\o'\imath_{X'}\l)\right)+\o\w\imath_X\left(d(\o'\imath_{X'}\l)\right)-
\o'\w\imath_{X'}\left(d(\o\imath_X\l)+\o\w\imath_Xd\l\right)\nonumber\\
&=\left(d\o\w(\imath_X\o')\imath_{X'}+\o\w (d\imath_X\o')\imath_{X'}
+\o\w((\imath_Xd\o')\imath_{X'} +\o\w\o' \imath_Xd\imath_{X'})\right.\nonumber\\
&-\left.\o'\w((\imath_{X'}d\o)\imath_X+(\imath_{X'}\o)\w d\imath_X+
\o\w\imath_{X'}d\imath_X+(\imath_{X'}\o)\w\imath_Xd+\o\imath_{X'}\imath_Xd)
\right)\l\qquad\nonumber\\[5pt]
&=\left(\imath_{\left[(d\o\w\imath_X \o'+\o\w{\cal L}_X\o')\otimes X'-\o'\w\imath_{X'}d\o\otimes X\right]}+\o\w\o' [{\cal L}_X,\imath_{X'}]-\o'\w\imath_{X'}\o\w{\cal L}_X\right)\l
\end{align}
where ${\cal L}_X=\imath_X d+d\imath_X$.
The last two terms can be rewritten using
\be [{\cal L}_X,\imath_{X'}]=\imath_{[X,X']}\ ,\ee
which follows by evaluating both sides on forms of degree zero and one, and
using the fact that if $\eta$ is an odd form then
\be \eta\w {\cal L}_X ={\cal L}_{\eta\otimes X}+d\eta\w\imath_X \ ,\ee
that we then apply for $\eta=\o'\w\imath_{X'}\o$. Thus,
\be[{\cal L}_{\omega\otimes X},\imath_{\omega'\otimes X'}]\lambda=
\imath_{\o\w\o'\otimes[X,X']+(d\o\w\imath_X \o'+\o\w{\cal L}_X\o')\otimes X'
-(\o'\w\imath_{X'}d\o+d(\o'\w\imath_{X'}\o))\otimes X}\l-
{\cal L}_{\o'\w\imath_{X'}\o \otimes X}\l\ ,\ee
in agreement with \eqref{Libracket} (again under the assumption that $\Xi$ is even).

\section{Covariant local cohomology of $\mathcal L_\mathcal Q$ on $T^\ast [1,0](T[0,1] M)$}
\label{App:C}

The construction of the supersymmetric covariant Hamiltonian
action makes use of differential forms ${\cal F}$ on $T^\ast [1,0](T[0,1] M)$ 
that are at least linear in momenta and annihilated by ${\cal L}_{\cal Q}$ where 
\be 
\mathcal Q := q_{\rm f} + \widetilde q_{\rm f}\ ,\qquad
q_{\rm f} =\theta^\a \tfrac{\partial}{\partial \phi^\a}\ ,\qquad
\widetilde q_{\rm f}=- \omega_\a  \tfrac{\partial}{\partial \chi_\a}\ ,\ee
is the nilpotent vector field on $T^\ast [1,0](T[0,1] M)$ of bi-degree $(0,1)$
that is the uplift of the de Rham differential on $M$.
Such forms are ${\cal L}_{\cal Q}$ exact and admit potentials 
that are ${\rm Diff}(M)$ covariant.
To demonstrate this, let us study the problem
\be {\cal L}_{\cal Q}{\cal F}={\cal J}\ ,\qquad {\cal L}_{\cal Q}{\cal J}=0\ .\ee
Introducing the nilpotent vector field
\be \overline{\mathcal Q} := \overline{\widetilde q}_{\rm f}\ ,\qquad
\overline{\widetilde q}_{\rm f}=-\chi_\a  \tfrac{\partial}{\partial \omega_\a}\ ,\ee
of ${\rm bideg}(\overline{\cal Q})=(0,-1)$, we can define the 
homotopy contracting vector field
\be
\mathcal{N}:=\{\mathcal{Q},\overline{\cal Q}\}= 
\omega_\a  \tfrac{\partial}{\omega_\a} +\chi_\a \tfrac{\partial}{\partial \chi_\a}\ ,
\ee
without breaking the ${\rm Diff}(M)$ covariance.
If we expand
\be \mathcal F=\sum_{p,q,r} \mathcal F_{[p|q,r]}\ ,\qquad {\rm deg}_{\cal E} (\mathcal F_{[p|q,r]})=p\ ,\qquad
{\rm bideg}(\mathcal F_{[p|q,r]})=(q,r)\ ,\ee
\emph{idem} ${\cal J}$, where thus
\be q\geqslant p\geqslant 0\ .\ee
then it follows from 
\be [\mathcal Q, \mathcal N] = 0=[\overline{\mathcal Q}, \mathcal N] \ ,\ee
and
\be {\cal L}_{\mathcal N} \mathcal F_{[p|q,r]}=q\mathcal F_{[p|q,r]}\ ,\label{QN}\ee
that we can fix the value of $q$.
Moreover, from the assumption on the the dependence of ${\cal F}$ on the momentum
variables $(\omega_\a,\chi_\a)$ it follows that
\be q\geqslant 1\ .\ee
Hence,
\be {\cal F}=({\cal L}_{\cal N})^{-1} {\cal L}_{\overline{\cal Q}} {\cal J}+ {\cal L}_{\cal Q} {\cal G}\ ,\ee
whose decomposition into definite quantum numbers read
\be {\cal F}_{[p|q,r]}= q^{-1}{\cal L}_{\overline{\cal Q}} {\cal J}_{[p|q,r+1]}+ 
{\cal L}_{\cal Q} {\cal G}_{[p|q,r-1]}\  ,\ee
and we can choose ${\cal G}$ such that
\be {\cal L}_{\overline{\cal Q}}{\cal G}=0\ .\ee

\paragraph{Zero-form ($p=0$).}
A function $\mathcal F$ on $T^\ast [1,0](T[0,1] M)$ that is at least 
linear in momenta can be expanded as
\be 
\mathcal F= \sum_{q \geqslant 1} \mathcal F_{(q)}\ ,\qquad 
 \mathcal F_{(q)}= \sum_{m+n=q} \mathcal F_{(m,n)}\ ,\qquad \mathcal F_{(m,n)} = \tfrac{1}{m! n!} \, \omega^m_{\a[m]} \,
                    \chi^n_{\b(n)} \, \mathcal F^{\a[m],\b(n)}_{(m,n)}~,
\ee
using a notation in which $\a[m]$ and $\b(n)$ stand for $m$ antisymmetric
and $n$ symmetric indices, respectively.
Thus, if $\mathcal Q  \mathcal F=0$, then $\mathcal F_{(q)}=\mathcal  Q \mathcal G_{(q)}$ 
where one can choose $\overline{\cal Q}{\cal G}_{(q)}=0$, which in particular
implies that $\mathcal G_{(q)}|_{\chi_\a=0}=0$.
It is instructive to arrive at this result by instead relying on
the local Poincar\`e lemma on $M$.
To this end, in the simplest case, one has  
\be
\label{F1}
\mathcal F_{(1)} = \mathcal F_{(1,0)} + \mathcal F_{(0,1)}
                 = \omega_\a \mathcal F^\a_{(1,0)}  + \chi_\a \mathcal F^\a_{(0,1)} .
\ee
It follows that 
\be
\label{QF1}
\mathcal Q \mathcal F_{(1)}=  -\omega_\a \, q_{\rm f}\mathcal F^\a_{(1,0)} 
                           + \chi_\a \, q_{\rm f}\mathcal {F}^\a_{(0,1)}  
                           + \omega_\a \mathcal {F}^\a_{(0,1)}
\ee
and thus, imposing $\mathcal Q \mathcal F_{(1)}=0$, one has the conditions
\be
\mathcal{F}^\a_{(0,1)} -q_{\rm f}\mathcal F^\a_{(1,0)} =0 ~~~,~~~ q_{\rm f}\mathcal {F}^\a_{(0,1)}=0,
\ee
whose general solution is given by
\be
\mathcal {F}^\a_{(0,1)}= q_{\rm f} \mathcal G^\a_{(0,1)} ~~~, 
~~~ \mathcal F^\a_{(1,0)}= \mathcal G^\a_{(0,1)} + q_{\rm f}\mathcal G^\a_{(1,0)} \ .
\ee
Plugging it back into \eqref{F1} one obtains
$\mathcal F_{(1)}=\mathcal Q \mathcal G_{(1)}$, with
\be
\label{G1}
\mathcal G_{(1)}= \omega_\a {\cal G}^\a_{(1,0)} +\chi_\a {\cal G}^\a_{(0,1)}\ .
\ee
The symmetry transformations
\be
\delta {\cal G}^\a_{(0,1)}=q_{\rm f} \Lambda^\a_{(0,1)}  \ ,\qquad  \delta {\cal G}^\a_{(1,0)} = \Lambda^\a_{(0,1)} 
+ q_{\rm f}\Lambda^\a_{(1,0)} ~,
\ee
can be used to set ${\cal G}^\a_{(1,0)}=0$, and one concludes that 
\be
\mathcal G_{(1)}=\chi_\a {\cal G}^\a_{(0,1)}\ ,\qquad  \overline{\cal Q}\mathcal G_{(1)}=0\ .
\ee 
Turning to the case of $q=2$, which concerns the Hamiltonian function, we insert
\be
\mathcal F_{(2)} = \tfrac{1}{2}\, \chi^2_{\b(2)} \mathcal F^{\b(2)}_{(0,2)} 
            + \omega_{\a} \chi_{\b} \mathcal F^{\a \b}_{(1,1)}
            +\tfrac{1}{2}\,\omega^2_{\a[2]} \mathcal F^{\a[2]}_{(2,0)}
\ee
into ${\cal Q}\mathcal F_{(2)}=0$, which yields the Cartan integrable system
\begin{align}
q_{\rm f}{\cal F}^{\a(2)}_{(0,2)}&=0\ ,\\[5pt]
q_{\rm f}{\cal F}^{\a,\b}_{(1,1)}+\mathcal F^{\a\b}_{(0,2)}&=0\ ,\\[5pt]
q_{\rm f}{\cal F}^{\b[2]}_{(2,0)}+2{\cal F}^{[\b,\b]}_{(1,1)} &=0\ ,
\end{align}
with general solution
\begin{align}
{\cal F}^{\a(2)}_{(0,2)}&=q_{\rm f} {\cal G}^{\a(2)}_{(0,2)}\ ,\\[5pt]
{\cal F}^{\a,\b}_{(1,1)}&=-q_{\rm f} {\cal G}^{\a,\b}_{(1,1)}-{\cal G}^{\a(2)}_{(0,2)}\ ,\\[5pt]
{\cal F}^{\b[2]}_{(2,0)}&=q_{\rm f} {\cal G}^{\b[2]}_{(2,0)}+2{\cal G}^{[\b,\b]}_{(1,1)}\ .
\end{align}
It follows that ${\cal F}_{(2)}={\cal Q}{\cal G}_{(2)}$, with
\be {\cal G}_{(2)}=\tfrac{1}{2}\, \chi_{\b(2)} \mathcal G^{\b(2)}_{(0,2)} 
            + \omega_{\a} \chi_{\b} \mathcal G^{\a \b}_{(1,1)}
            +\tfrac{1}{2}\,\omega_{\a[2]} \mathcal G^{\a[2]}_{(2,0)}\ ,\ee
modulo symmetry transformations.
Using the parameters of ${\cal F}^{\a(2)}_{(0,2)}$ and ${\cal F}^{[\a,\a]}_{(1,1)}$
to eliminate ${\cal G}^{(\a,\a)}_{(1,1)}$ and ${\cal G}^{[\b,\b]}_{(0,2)}$, respectively, we 
arrive at
\be {\cal G}_{(2)}=\tfrac{1}{2}\, \chi_{\b(2)} \mathcal G^{\b(2)}_{(0,2)} 
            + \omega_{\a} \chi_{\b} \mathcal G^{[\a \b]}_{(1,1)}\ ,\qquad\overline{\cal Q}{\cal G}_{(2)}=0\ .\ee
The general case follows the same pattern, such that if ${\cal Q}{\cal F}_{(q)}=0$ then 
we can write ${\cal F}_{(q)}={\cal Q}{\cal G}_{(q)}$ where
\be 
\mathcal G_{(q)}= \sum_{m=0}^{q-1} \tfrac{1}{m! (q-m)!} \, \omega^m_{\a[m]} \,
                    \chi^{q-m}_{\a\b(q-m-1)} \, \mathcal G^{[\a[m],\a]\b(q-m-1)}_{(m,q-m)}\ ,
                    \qquad\overline{\cal Q}{\cal G}_{(2)}=0\ . \ee
 %



\begin{thebibliography}{99}

\bibitem{Chu}
C.~Chu and P.~Ho,``\textit{Poisson Algebra Of Differential Forms}'', \href{http://dx.doi.org/10.1142/S0217751X97002929}{Int. J. Mod. Phys. 12 (1997) 5573-5587},
\href{http://arxiv.org/abs/q-alg/9612031}{arXiv:q-alg/9612031}.

\bibitem{Beggs}
E.~J.~Beggs and S.~Majid, ``\textit{Semiclassical differential structures}", \href{http://arxiv.org/abs/math/0306273}{arXiv:math/0306273}.

\bibitem{Tagliaferro}
A.~Tagliaferro, 
``\textit{A Star Product for Differential Forms on Symplectic Manifolds}", \href{http://arxiv.org/abs/0809.4717v2}{arXiv:0809.4717v2 [hep-th]}.

\bibitem{Zumino}
S.~McCurdy and B.~Zumino,
``\textit{Covariant Star Product for Exterior Differential Forms on Symplectic Manifolds}'',
\href{http://dx.doi.org/10.1063/1.3327559}{AIP Conf. Proc. 1200 (2010) 204-214},
\href{http://arxiv.org/abs/0910.0459}{arXiv:0910.0459 [hep-th]}.

\bibitem{Ikeda}
N.~Ikeda, ``\textit{Two-dimensional gravity and nonlinear gauge theory}", \href{http://dx.doi.org/10.1006/aphy.1994.1104}{Annals Phys. 235 (1994) 435-464}, \href{http://arxiv.org/abs/hep-th/9312059}{arXiv:hep-th/9312059}.

\bibitem{Schaller}
P.~Schaller and T.~Strobl, ``\textit{Poisson structure induced (topological) field theories}", \href{http://dx.doi.org/10.1142/S0217732394002951}{Mod. Phys. Lett. A9 (1994) 3129-3136}, \href{http://arxiv.org/abs/hep-th/9405110}{arXiv:hep-th/9405110}.

\bibitem{Cattaneo}
A.~S.~Cattaneo and G.~Felder, ``\textit{A Path integral approach to the Kontsevich quantization formula}", \href{http://dx.doi.org/10.1007/s002200000229}{Commun. Math. Phys. 212 (2000) 591-611}, \href{http://arxiv.org/abs/math/9902090}{arXiv:math/9902090}.

\bibitem{AKSZ}
M.~Alexandrov, M.~Kontsevich, A.~Schwartz and O.~Zaboronsky, ``\textit{The Geometry of the master equation and topological quantum field theory}", \href{http://dx.doi.org/10.1142/S0217751X97001031}{Int. J. Mod. Phys. A12 (1997) 1405-1430}, \href{http://arxiv.org/abs/hep-th/9502010}{arXiv:hep-th/9502010}.

\bibitem{Kontsevich}
M.~Kontsevich, ``\textit{Deformation quantization of Poisson manifolds}", \href{http://dx.doi.org/10.1023/B:MATH.0000027508.00421.bf}{Lett. Math. Phys. 66 (2003) 157-216}, \href{http://arxiv.org/abs/arXiv:q-alg/9709040}{arXiv:q-alg/9709040 [q-alg] }.

\bibitem{us}
C.~Arias, N.~Boulanger, P.~Sundell and A.~Torres-Gomez, 
``\textit{2D sigma models and differential Poisson algebras}", 
\href{http://dx.doi.org/10.1007/JHEP08(2015)095}{JHEP 1508 (2015) 095}, \href{http://arxiv.org/abs/arXiv:1503.05625}{arXiv:1503.05625 [hep-th]}.

\bibitem{Fedosov:1994zz} B.~V.~Fedosov, 
``\textit{A Simple geometrical construction of deformation quantization}", \href{http://projecteuclid.org/euclid.jdg/1214455536}{J. Diff. Geom. 40 (1994) no.2, 213-238}.

\bibitem{Moyal:1949sk}
  J.~E.~Moyal,
  ``\textit{Quantum mechanics as a statistical theory}",
  Proc.\ Cambridge Phil.\ Soc.\   45 (1949) 99.

\bibitem{Bayen:1977ha}
  F.~Bayen, M.~Flato, C.~Fronsdal, A.~Lichnerowicz and D.~Sternheimer,
  ``\textit{Deformation Theory and Quantization. 1. Deformations of Symplectic Structures}",
  Annals Phys.\  111 (1978) 61.
  
\bibitem{Bayen:1977hb}
  F.~Bayen, M.~Flato, C.~Fronsdal, A.~Lichnerowicz and D.~Sternheimer,
  ``\textit{Deformation Theory and Quantization. 2. Physical Applications}",
  Annals Phys.\  111 (1978) 111.
  
\bibitem{Zucchini}
R.~Zucchini, ``\textit{Gauging the Poisson sigma model}", \href{http://dx.doi.org/10.1088/1126-6708/2008/05/018}{JHEP 0805 (2008) 018}, \href{http://dx.doi.org/10.1088/1126-6708/2008/05/018}{arXiv:0801.0655 [hep-th]}.


\bibitem{us2} R.~Bonezzi, P.~Sundell and A.~Torres-Gomez, 
``\textit{2D Poisson Sigma Models with Gauged Vectorial Supersymmetry}", JHEP 1508 (2015) 047, 
\href{https://arxiv.org/abs/1505.04959}{arXiv:1505.04959 [hep-th]}.

\bibitem{Engquist:2005yt}
  J.~Engquist and P.~Sundell,
  ``\textit{Brane partons and singleton strings}",
  Nucl.\ Phys.\ B  752 (2006) 206,
  \href{https://arxiv.org/abs/hep-th/0508124#}{arXiv:hep-th/0508124}.

\bibitem{Engquist2} 
J.~Engquist, P.~Sundell and L.~Tamassia, 
``\textit{On Singleton Composites in Non-compact WZW Models}'',
\href{http://dx.doi.org/10.1088/1126-6708/2007/02/097}{JHEP 0702 (2007) 097}, \href{http://arxiv.org/pdf/hep-th/0701051.pdf}{arXiv:hep-th/0701051}.

\bibitem{Sundborg:2000wp}
  B.~Sundborg,
  ``\textit{Stringy gravity, interacting tensionless strings and massless higher spins}",
  Nucl.\ Phys.\ Proc.\ Suppl.\  102 (2001) 113, 
  \href{https://arxiv.org/abs/hep-th/0103247}{arXiv:hep-th/0103247}.

\bibitem{Sezgin:2002rt}
  E.~Sezgin and P.~Sundell,
  ``\textit{Massless higher spins and holography}",
  Nucl.\ Phys.\ B  644 (2002) 303,
   Erratum: [Nucl.\ Phys.\ B  660 (2003) 403],
  \href{https://arxiv.org/abs/hep-th/0205131}{arXiv:hep-th/0205131}.
  
\bibitem{Klebanov:2002ja}
  I.~R.~Klebanov and A.~M.~Polyakov,
  ``\textit{AdS dual of the critical O(N) vector model}",
  Phys.\ Lett.\ B  550 (2002) 213,
  \href{https://arxiv.org/abs/hep-th/0210114}{arXiv:hep-th/0210114}.
  
\bibitem{Vasiliev1988} M.~A.~Vasiliev, 
``\textit{Equations of Motion of Interacting Massless Fields of All Spins as a Free Differential Algebra}, \href{http://dx.doi.org/10.1016/0370-2693(88)91179-3}{Phys. Lett. B209 (1988) 491-497}

\bibitem{Vasiliev1990} M.~A.~Vasiliev, 
``\textit{Consistent equation for interacting gauge fields of all spins in (3+1)-dimensions}, \href{http://dx.doi.org/10.1016/0370-2693(90)91400-6}{Phys. Lett. B243 (1990) 378-382}.

\bibitem{Giombi:2010vg}
  S.~Giombi and X.~Yin,
  ``\textit{Higher Spins in AdS and Twistorial Holography}",
  JHEP 1104 (2011) 086
\href{https://arxiv.org/abs/1004.3736}{arXiv:1004.3736 [hep-th]}.

\bibitem{Colombo1} N.~Colombo and P.~Sundell, 
``\textit{Twistor space observables and quasi-amplitudes in 4D higher spin gravity}", \href{http://dx.doi.org/10.1007/JHEP11(2011)042}{JHEP 1111 (2011) 042}, \href{http://arxiv.org/abs/arXiv:1012.0813}{arXiv:1012.0813 [hep-th]}.

\bibitem{Colombo2} N.~Colombo and~P. Sundell, 
``\textit{Higher Spin Gravity Amplitudes From Zero-form Charges}", \href{http://arxiv.org/abs/arXiv:1208.3880}{arXiv:1208.3880 [hep-th]}.

\bibitem{Didenko1} V.~E.~Didenko and E.~D.~Skvortsov, 
``\textit{Exact higher-spin symmetry in CFT: all correlators in unbroken Vasiliev theory}", \href{http://dx.doi.org/10.1007/JHEP04(2013)158}{JHEP 1304 (2013) 158}, \href{http://arxiv.org/abs/arXiv:1210.7963}{arXiv:1210.7963 [hep-th]}.

\bibitem{Didenko2} V.~E.~Didenko, J.~Mei and E.~D.~Skvortsov, 
``\textit{Exact higher-spin symmetry in CFT: free fermion correlators from Vasiliev Theory}", \href{http://dx.doi.org/10.1103/PhysRevD.88.046011}{Phys. Rev. D88 (2013) 046011}, \href{http://arxiv.org/abs/arXiv:1301.4166}{arXiv:1301.4166 [hep-th]}.

\bibitem{Witten1} E.~Witten, ``\textit{Chern-Simons gauge theory as a string theory}", \href{http://dx.doi.org/10.1007/978-3-0348-9217-9_28}{Prog. Math. 133 (1995) 637-678}, \href{http://arxiv.org/abs/hep-th/9207094v2}{arXiv:hep-th/9207094v2}. 

\bibitem{Bershadsky} M.~Bershadsky, S.~Cecotti, H.~Ooguri and C.~Vafa,
``\textit{Kodaira-Spencer Theory of Gravity and Exact Results for Quantum String Amplitudes}"
\href{http://dx.doi.org/10.1007/978-3-0348-9217-9_28}{Commun. Math. Phys.165:311-428,1994}, \href{https://arxiv.org/abs/hep-th/9309140}{arXiv:hep-th/9309140}.

\bibitem{Marcus} N.~Marcus and A.~Sagnotti,
``\textit{Tree-level constraints on gauge groups for type I super-strings}"
\href{http://dx.doi.org/10.1007/978-3-0348-9217-9_28}{Phys. Lett. B 119, 97 (1982)}. 

\bibitem{GZ1} M.~R.~Gaberdiel and B.~Zwiebach,
``\textit{Tensor Constructions of Open String Theories I: Foundations}"
\href{http://dx.doi.org/10.1007/978-3-0348-9217-9_28}{Nucl. Phys. B505 (1997) 569-624}, \href{http://arxiv.org/abs/hep-th/9705038}{arXiv:hep-th/9705038}.

\bibitem{GZ2} M.~R.~Gaberdiel and B.~Zwiebach,
``\textit{Tensor Constructions of Open String Theories II: Vector bundles, D-branes and orientifold groups}"
\href{http://dx.doi.org/10.1007/978-3-0348-9217-9_28}{Phys. Lett. B410 (1997) 151-159}, \href{http://arxiv.org/abs/hep-th/9707051}{arXiv:hep-th/9707051}.

\bibitem{FCS} N.~Boulanger, E.~Sezgin and P.~Sundell, 
``\textit{4D Higher Spin Gravity with Dynamical Two-Form as a Frobenius--Chern--Simons Gauge Theory}",
\href{https://arxiv.org/abs/1505.04957v2}{arXiv:1505.04957v2 [hep-th]}.

\bibitem{FCS2} R.~Bonezzi, N.~Boulanger, E.~Sezgin and P.~Sundell, 
``\textit{Frobenius-Chern-Simons gauge theory}", to appear.

\bibitem{Park}
J.~S.~Park, ``\textit{Topological open p-branes}",
\href{http://arxiv.org/abs/hep-th/0012141v3}{arXiv:hep-th/0012141v3}.

\bibitem{Ikeda2012}
N.~Ikeda, ``\textit{Lectures on AKSZ Topological Field Theories for Physicists}", \href{https://arxiv.org/abs/1204.3714v3}{arXiv:1204.3714v3 [hep-th]}.

\bibitem{Schwarz:1978cn}
A.~S. Schwarz, ``\textit {The Partition Function of Degenerate Quadratic Functional
  and Ray-Singer Invariants}", Lett. Math. Phys. 2 (1978) 247--252.

\bibitem{Schwarz:1984wk}
A.~S. Schwarz and Y.~Tyupkin, ``\textit{Quantization of antisymmetric tensors and
  ray-singer torsion}",  Nucl. Phys. B242 (1984) 436--446.

\bibitem{Horowitz:1989km}
G.~T. Horowitz and M.~Srednicki, ``\textit{A Quantum Field Theoretic Description of
  Linking Numbers and Their Generalization}",  Commun. Math. Phys. 130 (1990) 83--94.

\bibitem{Wu:1990ci}
S.-Y. Wu, ``\textit{Topological Quantum Field Theories on Manifolds With a
  Boundary}", Commun. Math. Phys. 136 (1991) 157--168.

\end{thebibliography}
\end{document}